\documentclass[11pt,a4paper]{article}
\pdfoutput=1
\usepackage{jheppub}
\hypersetup{pdfencoding=unicode, bookmarksopen=true, bookmarksnumbered}
\usepackage[nomath]{lmodern}
\usepackage{microtype}
\usepackage{amsthm,amsbsy,amsfonts,mathrsfs,enumerate,float,wrapfig,amsmath}
\usepackage{multirow}
\usepackage{subfigure}
\usepackage{longtable}
\usepackage{adjustbox} 
\usepackage{physics}
\usepackage {dashrule}
\usepackage{tikz}
\newcommand{\be}{\begin{equation}}
\newcommand{\ee}{\end{equation}}

\title{Exploring new constraints on K\"ahler moduli space of 6d $\mathcal{N}=1$ Supergravity}

\author[a,b]{Hee-Cheol Kim}
\author[b]{and Cumrun Vafa}

\affiliation[a]{Department of Physics, POSTECH, Pohang 37673, Korea}
\affiliation[b]{Jefferson Physical Laboratory, Harvard University, Cambridge, MA 02138, USA}

\abstract{We propose new constraints for 6d $(1,0)$ supergravity theories based on consistency conditions on the K\"ahler moduli spaces of their 5d reductions. The requirement that both the metric and the BPS string tensions in the K\"ahler moduli space are positive imposes specific restrictions on the Chern-Simons coefficients in the 5d effective Lagrangians that are derived from the Kaluza-Klein reductions of 6d theories. Moreover, the emergence of local interacting 5d CFTs when the moduli space metric degenerates introduces additional constraints coming from the analysis of 5d SCFTs.  Focusing on the moduli spaces of 6d supergravity theories without a tensor multiplet and their Higgsings, we show that these constraints require the presence of certain primary states in the 2d worldvolume CFTs on 1/2 BPS strings. We specifically analyze a class of $SU(2)$ models and infinite families of $U(1)$ models using these constraints, and demonstrate that the theories featuring a 1-form symmetry in their massless spectra, unless the 1-form symmetry is gauged, fail to satisfy the constraints and therefore belong to the Swampland.}

\begin{document}

\maketitle

\section{Introduction}

Swampland program \cite{Vafa:2005ui} (see \cite{Brennan:2017rbf,Palti:2019pca,vanBeest:2021lhn,Grana:2021zvf,Agmon:2022thq} for reviews) provides a set of principles that aims to summarize what we have learned about allowed consistent quantum gravitational theories based on lesson learned from string theory landscape.  To assess the completeness of such principles, and also to see if the string theory landscape is the only consistent one, it would be interesting to see to what extent we can recover the full string landscape from such principles.

Theories with higher supersymmetry is a natural setting to start with this plan, as they are far more restrictive.  In fact successful applications of Swampland principles to theories with $N\geq 16$ supercharges shows the relative completeness of Swampland principles \cite{Adams:2010zy,Garcia-Etxebarria:2017crf,Kim:2019vuc,Kim:2019ths,Cvetic:2020kuw,Montero:2020icj,Dierigl:2020lai,Hamada:2021bbz,Bedroya:2021fbu}.
For theories with $N=8$ supercharges specifically starting in 6d with ${\cal N}=(1,0)$ supersymmetry, some progress in this direction has been made (See \cite{Sagnotti:1992qw,Kumar:2009ae,Kumar:2009ac,Kumar:2010ru,Kumar:2010am,Seiberg:2011dr,Kim:2019vuc,Lee:2019skh,Tarazi:2021duw,Baykara:2023plc,Hamada:2023zol,Becker:2023zyb,Loges:2024vpz,Hamada:2024oap}).  There is still some large discrepancy between what Swampland principles allow, and the string landscape with this much supersymmetry.  The aim of this paper is to try to provide some tools to narrow this gap, focusing on theories with small matter content in 6d:  Theories with no tensor multiplet and rank-1 gauge group, $SU(2)$ or $U(1)$.

The method we use involves computing the 5d effective Lagrangian for a 6d $(1,0)$ supergravity theory after a Kaluza-Klein (KK) circle reduction, and then testing the consistency of the Chern-Simons (CS) coefficients in this Lagrangian by exploring various corners in the Coulomb branch of the moduli space. We focus on testing consistency conditions derived from the positivity of the metric and that of BPS string tensions on the 5d Coulomb branch moduli space, as well as from the bounds on the cubic CS coefficients in local 5d superconformal field theories (SCFTs) that emerge at special loci on the moduli space. While this task might seem straightforward after the 5d effective Lagrangian is derived, as we will demonstrate, we find that checking these conditions is a highly non-trivial task. This process turns out to require correctly charting a specific region of the Coulomb branch, referred to as the {\it K\"ahler cone}. This region is dual to the cone of electric charges of 5d BPS particle states, and the positivity conditions and other criteria necessary for a unitary theory can be applied only in this region. 

We will introduce a systematic algorithm to identify the K\"ahler cones for the 5d theories derived from 6d supergravity theories. To do this, we analyze the spectrum of 5d BPS states that can arise from the chiral primary fields in 2d worldsheet CFTs living on BPS strings in the 6d theory, as well as those from the 6d massless matter fields and their KK momentum modes. The interplay between the charged primary fields and current algebras for the bulk gauge symmetries in the 2d CFTs will play an essential role in identifying certain non-perturbative BPS spectrum from 6d strings. This spectrum analysis will allow us to predict the potential boundaries of the K\"ahler cone within the Coulomb branch.

Based on this framework, we will investigate a large class of 6d supergravity theories without a tensor multiplet, including $SU(2)$ models and an infinite family of $U(1)$ models. In particular, we will prove that, among these, the theories with discrete 1-form symmetry in their massless spectra are inconsistent and belong to the Swampland unless the 1-form symmetry is promoted to a gauge symmetry of the full theory. This investigation provides robust field-theoretic evidence supporting the Massless Charge Sufficiency Conjecture recently proposed in \cite{Morrison:2021wuv} and the No Global Symmetries Conjecture in \cite{Banks:1988yz,Kallosh:1995hi,Banks:2010zn,Harlow:2018jwu,Harlow:2018tng,McNamara:2019rup,Harlow:2020bee}.

Even though we are not able to rule out many theories which look apparently consistent but not constructible in the string landscape, we nevertheless rule out some small subset of them.  More importantly we feel we have introduced some of the main ingredients that can be helpful in the future exploration of this topic.

The paper is organized as follows. Section \ref{sec:2} provides a review of the basic structure of 6d $(1,0)$ supergravity theories, including aspects of anomaly cancellation and 5d effective Lagrangians under a circle compactification. We then derive several constraints on CS coefficients in the 5d Lagrangian. Section \ref{sec:3} introduces a method to identify the K\"ahler cone of the 5d theory, which is based on the analysis of primitive BPS states originating from 6d massless multiplets and worldsheet primary fields on BPS strings. In Section \ref{sec:4}, we demonstrate how to apply this method to examine the consistency of anomaly-free theories with explicit examples of 6d $T=0$ models. A summary and some additional comments are given in Section \ref{sec:5}.

\section{K\"ahler moduli space of 6d $\mathcal{N}=1$ supergravity}\label{sec:2}

In this section, we discuss the compactification of 6d $\mathcal{N}=1$ supergravity theories to five dimensions. We particularly focus on the properties of the K\"ahler moduli space, which is parameterized by the real scalar fields coming from 6d tensor and vector multiplets. We start by reviewing basic features of generic 6d $\mathcal{N}=1$ supergravity theories. We then present a systematic computation of the 5d effective prepotential on the Coulomb branch of the moduli space in the 5d theory originating from the 6d gravity theory, and discuss the structure of the moduli space based on the characteristics of the effective prepotential.

Let us consider a 6d $(1,0)$ supergravity theory whose massless spectrum consists of a gravity multiplet, $T$ tensor multiplets, $V$ vector multiplets with gauge group $G$, and $H$ hypermultiplets. The massless spectrum is severely constrained by the cancellation of local and global anomalies with the help of the Green-Schwarz-Sagnotti mechanism \cite{Green:1984sg,Sagnotti:1992qw}. The conditions for the local gauge and gravitational anomaly cancellation can be summarized as
\begin{align}\label{eq:anomaly-cancel}
    &H - V = 273-29 T \ , \quad a\cdot a = 9-T \ , \nonumber \\
    & B_{\bf adj}^i = \sum_{\bf r}n^i_{\bf r}B^i_{\bf r} \ , \quad a\cdot b_i = \frac{\lambda_i}{6}\left(A^i_{\bf adj} - \sum_{\bf r}n_{\bf r}^iA_{\bf r}^i\right) \ , \nonumber \\
    &b_i\cdot b_i = \frac{\lambda_i^2}{3}\left(\sum_{\bf r}n^i_{\bf r}C^i_{\bf r}-C^i_{\bf adj}\right) \ , \quad b_i\cdot b_j = 2\lambda_i\lambda_j\sum_{\bf r,s}n_{\bf r,s}^{ij}A^i_{\bf r}A^j_{\bf s} \quad i\neq j \ ,
\end{align}
where $a^\alpha,b^\alpha_i$ are vectors, which we call anomaly coefficients, in the space $\mathbb{R}^{1,T}$, $n^i_{\bf r}$ denotes the number of hypermultiplets in the representation ${\bf r}$ for gauge group $G_i$, and we use the notation $v\cdot w=\Omega_{\alpha\beta}v^\alpha w^\beta$ to denote the inner product with respect to a natural symmetric metric $\Omega_{\alpha\beta}$ on this space. Here, $A_{\bf r}^i, B_{\bf r}^i, C_{\bf r}^i$ are the group theory factors defined as
\begin{align}
    {\rm tr}_{\bf r}F^2 = A_{\bf r}{\rm tr}F^2 \ , \quad {\rm tr}_{\bf r}F^4=B_{\bf r}{\rm tr}F^4 + C_{\bf r}\left({\rm tr}F^2\right)^2 \ ,
\end{align}
and $\lambda_i$ is another group theory factor that normalizes the smallest topological charge of a unit instanton for a group $G_i$ to be `1' which is summarized, for example, in \cite{Kumar:2010ru}.

The $(1,0)$ tensor multiplets contain real scalar fields, and their vacuum expectation values, which can be collected into a vector $J^\alpha$ in the space $\mathbb{R}^{1,T}$, parametrize the tensor branch of a K\"ahler moduli space defined by the coset space $SO(1,T)/SO(T)$. We impose the conditions that the vector $J^\alpha$ satisfies $J\cdot J =1$ to ensure that the gravitational coupling remains finite, and $J\cdot b_i > 0$ to maintain the positivity of the gauge kinetic terms. In F-theory models, this vector $J^\alpha$ corresponds to the K\"ahler form of the base 4-manifold in the associated Calabi-Yau 3-fold.

\subsection{Circle compactifications}

Let us move on to the circle reduction of a 6d supergravity theory to five dimensions, which results in a 5d $\mathcal{N}=1$ supergravity theory at low energy. In this reduction, the gravity multiplet in the 6d reduces to a 5d gravity multiplet, which includes a gravi-photon vector field, and a 5d vector multiplet. Similarly, the 6d hyper and vector multiplets become 5d hyper and vector multiplets respectively. Additionally, a 6d tensor multiplet gives rise to a 5d vector multiplet. 

We can turn on gauge holonomies along the 6d circle, and they provide real scalar fields in the 5d vector multiplets. The collection of the scalar expectation values in these 5d vector multiplets forms the Coulomb branch of the moduli space, denoted by $\mathcal{K}$. The rank (or the dimension) of the Coulomb branch moduli spaces is therefore given by ${\rm rank}(\mathcal{K}) = T + {\rm rank}(G) + 1$, where $G$ denotes the 6d gauge groups and $+1$ accounts for the vector multiplet from the 6d gravity multiplet.

The low energy dynamics on the Coulomb branch of a 5d theory is characterized by the prepotential $\mathcal{F}$ which is a cubic polynomial in the scalar expectation values in the 5d vector multiplets, which we denote by homogeneous coordinates $t_I$ with $I=0,1,\cdots,{\rm rank}(\mathcal{K})$ subject to the equivalence relation $t_I \cong \lambda t_I$. The cubic Chern-Simons Lagrangian at a generic point on the Coulomb branch is determined from the prepotential as
\begin{align}
    \mathcal{L}_{CS} = - \frac{1}{6} C_{IJK}A^I \wedge F^J\wedge F^K \ , \quad C_{IJK} = \partial_I\partial_J\partial_K \mathcal{F}(t) \ ,
\end{align}
with Chern-Simons coefficients $C_{IJK} \in \mathbb{Z}$, where $\partial_I \equiv \partial/\partial t_I$. The gauge kinetic matrix $M_{IJ}$ in the low energy theory is also determined by the prepotential as
\begin{align}
    M_{IJ} = \partial_I \mathcal{F}\partial_J \mathcal{F} - \partial_I \partial_J \mathcal{F}\ .
\end{align}
This also defines the metric on the Coulomb branch defined as a hypersurfaces of $t_I$'s under the condition $\mathcal{F}=1$. We remark that the gauge kinetic matrix must be positive (semi-)definite in the moduli space of a consistent 5d supergravity theory.

The 5d effective action for a circle compactification of a generic 6d $\mathcal{N}=1$ supergravity theory is computed in \cite{Bonetti:2011mw,Bonetti:2013ela,Grimm:2015zea}. This computation was carried out by taking into account the contributions from the 6d classical actions on the tensor branch including the Green-Schwarz terms and also the one-loop contributions which are induced by integrating out the massive fields after the circle compactification. Here, the one-loop contributions also involve contributions from the Kaluza-Klein modes that couple to the 5d gravi-photon field. With all these factors considered, the 5d effective Chern-Simons Lagrangian resulting from the circle reduction can be expressed as \cite{Bonetti:2011mw,Bonetti:2013ela,Grimm:2015zea}
\begin{align}\label{eq:Lagrangian}
    \mathcal{L}_{CS}^{5d} &= -\frac{9-T}{24}A^0\wedge F^0\wedge F^0 - \frac{1}{2}A^0\wedge (\Omega_{\alpha\beta}F^\alpha\wedge F^\beta) \nonumber \\
    & \quad +\frac{1}{4}\sum_ib_{\alpha}^i\cdot \left(A^\alpha\!-\!\tfrac{1}{2}a^\alpha A^0\right)\wedge {\rm tr}(F^i\wedge F^i)  + \mathcal{L}_{\rm z.m} \ ,
\end{align}
where $A^0$ is the Kaluza-Klein vector field, $F^\alpha=dA^\alpha$ denotes the field strength for $A^\alpha$ with $\alpha=1,\cdots,T+1$ obtained by the circle reduction of the 6d 2-form tensor field $B^\alpha$, and $F^i = dA^i$ is the field strength for the gauge group $G_i$. The term $\mathcal{L}_{\rm z.m}$ stands for the contributions from the 6d vector- and hyper-multiplets with zero KK charge. When gauge holonomies along the 6d circle, which become scalar expectation values $\phi^i$ in 5d, are turned on, this term can be written as \cite{Seiberg:1996bd,Intriligator:1997pq}
\begin{align}\label{eq:pre-pert}
    \mathcal{L}_{\rm z.m} &= -\frac{1}{6} C_{abc}^{\rm z.m}A^a\wedge F^b \wedge F^c \ , \quad C_{abc}^{\rm z.m} = \partial_a\partial_b\partial_c \mathcal{F}_{\rm z.m}(\phi) \ , \nonumber \\
    \mathcal{F}_{\rm z.m} &= -\frac{1}{12}\left(\sum_{e\in {\bf R}}|e\cdot\phi|^3 -\sum_{f}\sum_{w \in {\bf w}_f} |w(\phi)|^3\right) \ ,
\end{align}
where ${\bf R}$ are the roots of the 6d gauge group $G$, ${\bf w}_f$ are the weights of the representation ${\bf r}_f$ of the hypermultiplets, and the indices $a,b,c$ here run over Cartan elements of the 6d gauge group.

\subsection{Constraints on the moduli space}

Now we discuss the constraints on the Coulomb branch moduli space $\mathcal{K}$ based on the effective prepotential in the low-energy theories of 5d gravity. This subsection is dedicated to the constraints applicable to any generic 5d supergravity theory. Later, in the next section, we will discuss additional constraints specific to 5d theories that arise from circle reductions of 6d supergravity theories.

Let us first introduce some basic terms related to the Coulomb branch of the moduli space. In a 5d supergravity theory, the BPS states on the Coulomb branch include electrically charged particles and their dual magnetic monopole strings. The masses and tensions of these BPS states are non-negative within the physical region of the Coulomb branch moduli space. This physical region is represented as a cone in a real space whose dimension equals to the rank of $\mathcal{K}$. We shall refer to this cone of the physical Coulomb branch as a {\it K\"ahler cone}. The boundary of the K\"ahler cone is defined as a region where certain BPS particle states become massless. We can therefore define the K\"ahler cone as the dual cone of the cone of BPS particle states. This cone of BPS particle states is often called Mori cone in geometric models constructed by M-theory compactification on a Calabi-Yau 3-fold.

The cone of BPS particle states is generated by primitive BPS particle states, whose masses vanish at a boundary of the dual K\"ahler cone. Here, a primitive state is defined as one whose electric charges cannot be expressed as a positive linear combination of charges of other BPS states. Given that the K\"ahler cone is a dual cone, the masses of these primitive states, which can be written as $M \sim q_it_i$ for a state with electric charge $q_i$, serve as the generators of the dual K\"ahler cone. The cone of BPS particle states and its dual K\"ahler cone are referred to as {\it simplicial} when the number of their generators equals ${\rm rank}(\mathcal{K})+1$. If there are more generators than this, the cones are called {\it non-simplicial}. 

We can write the prepotential in terms of the K\"ahler cone generators $t_I$ as
\begin{align}
    \mathcal{F}(t)=\frac{1}{6} C_{IJK} t^I t^J t^K \ .
\end{align}
Let us first consider the cases when the K\"ahler cone is simplicial. In such cases, the K\"ahler cone precisely agrees with the cone $\mathcal{C}(t_I)$ generated by $t_I$ for $I=0,1,\cdots, {\rm rank}(\mathcal{K})$. This means that $\mathcal{K}=\mathcal{C}(t_I)$. Then, this allows us to immediately derive several positivity constraints on the Chern-Simons coefficients $C_{IJK}$.

First, as mentioned above, the gauge kinetic matrix must be positive (semi-)definite $M_{IJ}\ge0$ within the K\"ahler cone $\mathcal{K}$ that is defined as a cone with $t_I\ge0$. Additionally, by imposing the positivity of BPS monopole tensions, which can be obtained by $T_I = \partial_I \mathcal{F}$, inside the K\"ahler cone, we find positivity constraints on the Chern-Simons coefficients as
\begin{align}\label{eq:CScoeff-positivity}
   T_I\ge 0 \quad \rightarrow \quad C_{III} \ge0 \ ,\quad C_{IJJ} \ge 0 \quad J\neq I \ .
\end{align}
Here, the first condition arises from the requirement of positive tension in regions where all $t_J\rightarrow0$ with $J\neq I$, and the second condition is derived from the tension positivity around regions where both $t_I$ and $t_K\rightarrow0$ with $K\neq J$, while $t_J$ remains finite, such that $T_I\sim C_{IJJ}$.
These constraints become crucial in later sections when we examine the consistency of 6d theories through the prepotentials derived from their 5d reductions.

On the other hand, if the cone is non-simplicial, the K\"ahler cone $\mathcal{K}$ and therefore the physical region of the moduli space can be smaller than the cone $\mathcal{C}(t_I)$ formed by a certain combination of $t_I$'s with $I=0,\cdots,{\rm rank}(\mathcal{K})$. This means that $\mathcal{K}\subset \mathcal{C}(t_I)$. When this is the case, some region parametrized by these $t_I$ values  may lie outside the K\"ahler cone and thus the above positivity constraints on CS coefficients in the prepotential, which is written in terms of $t_I$, no longer hold. However, we can choose an alternative combination, say $\tilde{t}_I$'s with $I=0,\cdots,{\rm rank}(\mathcal{K})$, such that the cone $\mathcal{C}(\tilde{t}_I)$ with $\tilde{t}_I\ge0$ is strictly contained within the K\"ahler cone, i.e. $\mathcal{C}(\tilde{t}_I) \subset \mathcal{K}$.  The positivity constraints are then applicable within this smaller cone $\mathcal{C}(\tilde{t}_I)$. Therefore, when the Coulomb branch of a theory is non-simplicial, a more careful application of positivity conditions is needed. This distinction will be crucial in later discussions where we use moduli space analysis to investigate certain anomaly-free theories.

Additional constraints can be found by a more detailed analysis of the boundaries of K\"ahler cones where some Coulomb branch parameters $t_I$ vanish. Systematic studies of boundary behaviors of the Coulomb branch moduli space in 5d supergravity theories have been conducted in \cite{Alim:2021vhs,Rudelius:2023odg}, which we summarize now. First, a boundary can be located at an infinite distance in the moduli space. In such cases, the Distance Conjecture in \cite{Ooguri:2006in} implies that there must be a tower of particle states whose masses become exponentially light. Moreover, according to the Emergent String Conjecture in \cite{Lee:2019wij,Lee:2019xtm}, this tower must be either of a Kaluza-Klein tower, corresponding a decompactification limit, or of string oscillator modes, corresponding to an emergent string limit.
Specifically in 5d, such infinite distance limits can be characterized by asymptotic behaviors of the prepotential $\mathcal{F}$. Let us consider a straight path in homogeneous coordinates defined as $t_I(s) = t^I_0 + s\, t^I_1$ with a constant vector $t^I_0$, where the infinity is approached as $s \rightarrow 0$ (before forcing the constraint $\mathcal{F}=1$). As we approach to infinity with $s\rightarrow 0$, the prepotential is expected to vanish as follows: \cite{Etheredge:2022opl,Rudelius:2023odg}
\begin{align}
    &\bullet \ {\rm Decompactification \ limit} \ : \ \mathcal{F}\sim s \nonumber \\
    &\bullet\ {\rm Emergent \ string \ limit} \ : \ \mathcal{F}\sim s^2 \nonumber 
\end{align}
Thus, the CS coefficients in the prepotential control the infinite distance limits on the Coulomb branch moduli space.

On the other hand, a boundary of the K\"ahler cone can occur at finite distance. There are three possible scenarios in such cases: 1) an infinite number of light degrees of freedom at the finite distance boundary, 2) hypermultiplets becoming massless, and 3) vector multiplets (possibly along with hypermultiplets) becoming massless. The first scenario can arise when some eigenvalues of the metric $M_{IJ}$ vanish and thus the associated gauge coupling diverges \footnote{This is also the case where the scalar curvature of the vector multiplet moduli space in the 4d reductions diverges positively as discussed in \cite{Marchesano:2023thx}. Such divergence signals the presence of a CFT fixed point at the location of the divergence.}, as noted in \cite{Etheredge:2022opl,Rudelius:2023odg}. This thus implies the existence of a strongly interacting CFT fixed point localized at the finite distance boundary of the moduli space. The second case comes with a {\it flop transition}, which brings us to another chamber in the moduli space. The third scenario signals an enhancement of $SU(2)$ gauge symmetry, and an Weyl reflection takes place as we move further along the path. In summary, the finite distance boundaries of the K\"ahler cone, where some of the $t_I$'s vanish while $\mathcal{F}$ remains finite, are characterized as
\begin{align}
    \bullet\ {\rm CFT} \ : \ {\rm det}M_{IJ} = 0 \ ,\qquad  \bullet\ {\rm Flop\ or\ Weyl\ reflection} \ : \ {\rm Otherwise}  \nonumber 
\end{align}
The first case, where we encounter a local 5d CFT fixed point at the boundary, is particularly interesting in the subsequent discussions. In this case, the classifications of 5d SCFTs, as detailed in references like \cite{Seiberg:1996bd,Morrison:1996xf,Intriligator:1997pq,Jefferson:2017ahm,Jefferson:2018irk,Bhardwaj:2019jtr}, allow us to impose strong constraints on the CS coefficients.  We will further explore this aspect using rank-1 supergravity theories as concrete examples.

\subsection{Rank-1 supergravities}

5d rank-1 supergravity theories have one-dimensional Coulomb branch of the moduli space $\mathcal{K}$. Suppose that the K\"ahler cone of a 5d theory is parametrized by two generators $t_1\ge 0$ and $t_2\ge0$. The prepotential for any such theory can be written as
\begin{align}
    6\mathcal{F} = c_1\,t_1^3 + 3c_2\, t_1^2 t_2 + 3c_3\,t_1t_2^2 + c_4\, t_2^3 \ ,
\end{align} 
with four Chern-Simons coefficients $c_{1,2,3,4} \in \mathbb{Z}_{\ge0}$. As explained above, these CS coefficients are constrained to ensure that the prepotential corresponds to that of a consistent supergravity theory.  First, the metric positivity within $\mathcal{K}$ requires
\begin{align}
    \Delta_1 \equiv c_2^2-c_1c_3 \ge0 \ , \quad \Delta_2 \equiv c_3^2 - c_2c_4\ge 0 \ . 
\end{align}
We note that the same two conditions for rank-1 theories were also derived from an analysis of the consistency in current algebras of a worldsheet CFT on BPS monopole strings in \cite{Katz:2020ewz}.  These conditions suggest that once the coefficients in the cross terms are fixed, the number of possible prepotentials is limited.

If we know what happens at the boundaries, such as at $t_1=0$ or $t_2=0$, of the K\"ahler cone, more constraints can be extracted. Conversely, the structure of CS coefficients can reveal the dynamics occurring at these boundaries. We can summarize the relations between CS coefficients and the boundary physics near $t_2=0$ as follows:
\begin{itemize}
    \item $c_1,c_2=0, \ c_3 > 0$ \ : \ a tower of string oscillator modes (an emergent string limit)
    \item $c_1=0, \ c_2>0$ \ : \ a tower of KK states (a decompactification limit)
    \item $c_1>0$ and $\Delta_1=0$ \ : \ CFT fixed point
    \item $c_1>0$ and $\Delta_1>0$ \ : \ Flop transition or Weyl reflection
\end{itemize}
The relations for the boundary near $t_1=0$ can be determined similarly. As previously mentioned, the first two cases correspond to infinite distance limits, while the latter two correspond to finite distance boundaries. Also, in the first three cases there is an infinite number of BPS states with masses proportional to $nt_2$ for $n\in\mathbb{Z}_{>0}$. As we approach the boundary where $t_2\rightarrow0$, these states become massless. However, in the last case, where we have flop transitions or Weyl reflections, only a finite number of BPS states become massless at the boundary, and their masses are bounded by $nt_2$ for $0<n<r$, where $r$ is some finite number.

The condition $\Delta_1=0$ for a CFT boundary is from the fact that an eigenvalue of the metric $M_{IJ}$ vanishes at the boundary. The CFT when this is the case is a 5d rank-1 SCFT. The classification of 5d rank-1 SCFTs has been completed. There are only 11 rank-1 SCFTs: ten are geometric theories associated with local CY threefolds embedding a del Pezzo surface ${\rm dP}_n$ with $n=0,1,\cdots 8$ or a Hirzebruch  surface $\mathbb{F}_0$, often referred to as $E_n$, $\tilde{E}_1$ theories, as studied in \cite{Seiberg:1996bd,Morrison:1996xf,Intriligator:1997pq}, and one non-geometric theory predicted in \cite{Bhardwaj:2019jtr}. These rank-1 SCFTs have prepotentials of the form $6\mathcal{F}= (9-n)t^3$ for geometric $E_n$ (or $\tilde{E}_1$ with $n=1$) theories, or $6\mathcal{F}= t^3$ for the non-geometric one.\footnote{The $E_8$ theory and the non-geometric theory both have the same cubic prepotential $6\mathcal{F}= t^3$. These theories are distinguished by their second Chern class numbers (or the gauge-gravity mixed CS coefficients), which are $+10$ and $-2$ respectively.} This classification imposes a strong constraint on the cubic prepotential forms for a 5d theory that contains CFT fixed points on its Coulomb branch.


Let us assume that the metric on the Coulomb branch satisfies $\Delta_1=0$ near $t_2\rightarrow 0$ and thus we have a CFT fixed point at the boundary. The null vector $v^I$ of $M_{IJ}$, whose eigenvalue vanishes at the boundary (i.e. $M_{IJ}v^J = 0$), specifies the $U(1)$ gauge field that couples to the BPS states in the CFT at the fixed point. Specifically, all local BPS states of the CFT near the boundary carry electric charges for the $U(1)$ gauge field orthogonal to the null vector, $A_{I\mu}  v^I=0$. Then the existence of the local CFT forces that the cubic CS coefficient $C_{\rm cft}$ for this $U(1)$ gauge field must be within the range of $1\le C_{\rm cft}\le 9$, which is the range of the possible cubic CS coefficients for the rank-1 5d SCFTs. This is because the $U(1)$ gauge field becomes the vector field for the Coulomb branch of the rank-1 CFT at the boundary. If not, the prepotential with $\Delta_1=0$ at a boundary $t_2\rightarrow 0$ cannot support a CFT fixed point at the point, which results in inconsistency. Therefore, any 5d rank-1 supergravity theory with such a prepotential that has 
\begin{align}
    \Delta_1=0\ ({\rm or} \ \Delta_2=0) \quad {\rm and} \quad C_{\rm cft}>9 \qquad \rightarrow \quad {\it Swampland}
\end{align}
at a boundary $t_2=0$ (or $t_1=0$) is not a consistent theory and thus falls into the Swampland.

In this argument, it is crucial to use the correct quantization condition for the $U(1)$ gauge field. We quantize the $U(1)$ gauge field such that the minimal magnetic charge for a BPS monopole string with respect to the gauge field is `1'. This quantization condition is also related to the 1-form global symmetries for the CFTs localized at the boundary of the Coulomb branch. For instance, the $E_0$ CFT has a cubic CS coefficient `9' for the $U(1)$ gauge field under this quantization condition, and the BPS states in this CFT carry electric charges of $3n$ for $n\in\mathbb{Z}$. This leads to a $\mathbb{Z}_3$ 1-form global symmetry. However, this 1-form symmetry will be explicitly broken when the CFT is coupled to gravity. We will look at explicit examples of this.

As an example, let us consider a 5d rank-1 supergravity arising from M-theory compactified on an elliptic threefold fibered over $\mathbb{P}^2$. This theory also corresponds to a circle compactification of the 6d $(1,0)$ supergravity theory without tensor and vector multiplets. The triple intersection numbers for the Nef divisors $H$ and $L$ in the 3-fold were computed in \cite{Candelas:1994hw}, and in terms of their K\"ahler parameters, $t_0$ and $t_1$ respectively, the cubic prepotential for the 5d theory can be written as 
\begin{align}\label{eq:ellipticP2}
    6\mathcal{F}_{{\rm ell}\,\mathbb{P}^2} = 9t_0^3+9t_0^2 t_1 + 3 t_0t_1^2 \ .
\end{align}

There are two boundaries: one at $t_0=0$ and the other at $t_1=0$. The first boundary, where $t_0\sim 0$, is located at infinite distance. Since the prepotential vanishes as  $\mathcal{F}\sim t_0$ near this boundary, we expect a tower of light KK states that leads to a decompactification as we approach this boundary. Indeed, this theory originates from a 6d gravity theory on a circle, and the limit $t_0\rightarrow 0$ corresponds to the decompactification of the KK circle back to 6d.

The second boundary at $t_1\sim 0$ is located at a finite distance, as $\mathcal{F}$ remains finite around this region. Moreover, the CS coefficients satisfy $\Delta_1 = 3^2 - 9 = 0$, and the metric exhibits a vanishing eigenvector $v^I=(1,-3)^T$ near the boundary. This indicates the presence of a localized CFT at the boundary. Hence, we expect at this boundary that 2-cycles and a divisor associated with the K\"ahler parameter $\tilde{t}_1$ shrink at the same time, where $\tilde{t}_1$ is dual to the null vector $v^I$, defined by $t_0=\tilde{t}_0+\tilde{t}_1$ and $t_1 = -3\tilde{t}_1$. The Coulomb branch of the local CFT is now parameterized by the new K\"ahler parameter $\tilde{t}_1$. The cubic prepotential in terms of the K\"ahler parameters $\tilde{t}_0, \tilde{t}_1$ can be written as
\begin{align}\label{eq:localP2}
    6\mathcal{F}_{{\rm ell}\, \mathbb{P}^2} = 9\tilde{t}_0^3 + 9\tilde{t}_1^3\ .
\end{align}

One then find that the cubic CS coefficient for the local CFT, which is the coefficient of $\tilde{t}_1^3$ term, is `9'. This suggests that the CFT is a rank-1 $E_0$ CFT, which can be realized through the compactification of M-theory on a local Calabi-Yau (CY) 3-fold $\mathcal{O}(-3)$ over $\mathbb{P}^2$. This is actually expected. The limit $t_1\rightarrow 0$ corresponds to a corner in the moduli space where the size of the elliptic fibration in the elliptic CY 3-fold is taken to infinity, while the volume of the base $\mathbb{P}^2$ is kept finite. This results in the local CY 3-fold $\mathcal{O}(-3)$ over $\mathbb{P}^2$. 

Holomorphic 2-cycles in this geometry, which corresponds to BPS particle states in the CFT, have volumes given by $nt_1 = -3n\tilde{t}_1$ for $n\in \mathbb{Z}_{>0}$. This means that $\tilde{t}_1$ is negative on the physical region of the Coulomb branch. The negative sign in the volume formula, compared to that of the local Calabi-Yau geometry, reflects the fact that the divisor for the CFT, whose K\"ahler parameter is $\tilde{t}_1$, is an anti-effective Nef divisor \cite{Katz:2020ewz}. We also note that the minimal electric charge of BPS states in the local CFT with respect to $\tilde{t}_1$ is `3'. This implies that there is a $\mathbb{Z}_3$ 1-form global symmetry in the local theory. This local 1-form symmetry is however broken when the CFT is embedded in the gravity theory due to the presence of other massive BPS states carrying charge $+1$ for the $U(1)$ symmetry with mass $t_0=\tilde{t}_0 + \tilde{t}_1$.

We remark here that the same cubic prepotential as seen in \eqref{eq:ellipticP2} can also be derived directly from the effective Lagrangian \eqref{eq:Lagrangian} of the 6d gravity theory without using geometric data. For the 6d theory without tensor and vector multiplets, we have
\begin{align}
    \Omega = 1 \ , \quad a = -3 \ , \quad b_i = 0 \ ,
\end{align}
and thus the prepotential is
\begin{align}
    6\mathcal{F} = \frac{9}{4}t_0^3+3t_0 t_1^2 \ .
\end{align}
However, the parameters $t_0$ and $t_1$ used here are not the correct K\"ahler cone generators due to the KK charge shifts in the BPS string spectrum, which will be discussed more comprehensively in the next section. The correct K\"ahler cone generators are obtained by shifting $t_1 \rightarrow t_1 + \frac{3}{2} t_0$. With this correction, we can precisely reproduce the prepotential given in \eqref{eq:ellipticP2} derived from the geometric calculation. This analysis demonstrates that, even without relying on geometric data, we can solidly derive the fact that the 6d theory without tensor and vector multiplet contains the 5d $E_0$ SCFT at a finite distance in the moduli space of its 5d reduction. Moreover, this ensures the existence of BPS strings at every tensor charge in any $T=0$ supegravity theory since such a theory can always be Higgsed to the $T=0$ theory without gauge symmetry and after a circle reduction, the ground state in the worldsheet CFT on each wrapped string is mapped to a BPS state in the 5d $E_0$ theory. Therefore, the tensor charge lattice of any $T=0$ theory is completely populated by BPS strings.

\subsection{Higgsing}\label{sec:higgsing}

Analyzing the Coulomb branch moduli space becomes increasingly complex for higher-rank theories. To simplify our studies, we will Higgs these higher-rank theories to lower-rank ones and analyze the moduli space of the resulting lower-rank theories instead. While the moduli space after Higgsing only captures a subset of the full moduli space of the UV theory, it still allows us to extract some interesting constraints on the original theory. In the following discussions, we will consider 6d supergravity theories with $T=0$ and a rank-1 gauge algebra, such as $U(1)$ or $SU(2)$. These theories lead to rank-2 gravity theories in 5d under a circle reduction. We will Higgs these theories to rank-1 theories and investigate their moduli space. We leave further studies on higher-rank theories for future research.

Let us demonstrate the idea of Higgsing through a specific example. We consider a 6d supergravity theory with no tensor multiplet $(T=0)$ and an $SU(2)$ gauge algebra, coupled to $n_{\bf 3}$ adjoint and $n_{\bf 2}$ fundamental hypermultiplets. The anomaly cancellation with $a=-3$ and $1\le b\le 12$ requires the number of massless hypermultiplets to be \cite{Kumar:2010am}
\begin{align}\label{eq:su2-hyper}
    n_{\bf 2} = 2b(12-b) \ , \quad n_{\bf 3} = \frac{(b-1)(b-2)}{2} \ ,
\end{align}
together with $n_{\bf 1}$, the number of neutral hypermultiplets determined by the cancellation of gravitational anomaly.

These theories with $1\le b \le 8$ have known realizations in F-theory, and corresponding elliptic Calabi-Yau 3-folds can be constructed as toric hypersurfaces \cite{CYdata,Huang:2018vup,Hayashi:2023hqa}. Also, the theory with $b=12$, which has no fundamental hypermultiplet, can be realized in F-theory if the gauge group is $SO(3)$ rather than $SU(2)$ \cite{Morrison:2021wuv}. On the other hand, the F-theory backgrounds realizing the theories with $b=9$ and $b=10, 11$ exhibit enhanced gauge symmetries to $(SU(2)\times U(1))/\mathbb{Z}_2$ and  $(SU(2)\times SU(2))/\mathbb{Z}_2$, respectively, and they cannot be Higgsed to the theories with only an $SU(2)$ gauge symmetry \cite{Raghuram:2020vxm}. We will further investigate these theories with $9\le b\le 12$ in Section \ref{sec:su2}.

 The cubic prepotential for these theories after a circle reduction is written as \cite{Huang:2018vup,Hayashi:2023hqa}
\begin{align}\label{eq:su2-adj}
    6\mathcal{F}_{SU(2)_b} =&\ 9 t_0^3 + 9t_0^2 t_1 + 3 t_0t_1^2 +54 t_0^2 t_2 + 6 t_1^2 t_2 +36t_0t_1t_2 \nonumber \\
    & +18(6-b)t_0t_2^2 + 6(6-b)t_1t_2^2 + 2(6-b)^2 t_2^3 \ ,
\end{align}
in terms of three K\"ahler cone generators $t_{0,1,2}$. Here, $t_0$ is the K\"ahler parameter dual to the 5d state with KK charge $+1$ and 6d $U(1)\subset SU(2)$ gauge charge $-2$, and $t_1$ and $t_2$ are K\"ahler parameters for the tensor charge and the $U(1)$ gauge charge, respectively. One finds that some coefficients turn negative when $b>6$. This implies that the K\"ahler cone is non-simplicial in these cases. We will discuss the non-simplicial examples with $b=7,8$ in Section \ref{sec:chiralprimary}.

We can Higgs this theory by giving vacuum expectation values (vevs) to hypermultiplets and then consider the low energy limit. The Higgsings can be implemented at the level of the prepotential by turning off some K\"ahler parameters. For example, we can give vevs to fundamental hypermultiplets, except $b=12$ case. Then, this leads to the theory without tensor and vector multiplets that we previously discussed at low energy. This Higgsing can be realized by setting $t_2=0$ in the prepotential. One can see that the prepotential with $t_2=0$ indeed reduces to that of the elliptic $\mathbb{P}^2$ theory given in \eqref{eq:ellipticP2}, as expected.

A more non-trivial Higgsing process is possible for the theories with $b\ge 4$ when compactified on a circle. These theories contain multiple adjoint hypermultiplets, and they can be used for a Higgsing process. We first give a vev to an adjoint hypermultiplet. This breaks the $SU(2)$ gauge symmetry to $U(1)$. The resulting theory has hypermultiplets with electric charge $\pm2$ as well as those with charge $\pm1$. After a circle compactification, the BPS spectrum of the 5d theory includes a hypermultiplet state with electric charge $-2$ and KK charge $+1$ along the 6d circle. The 5d theory can be further Higgsed by giving a vev to this particular hypermultiplet, which sets a K\"ahler parameter to zero, $t_0=0$. The resulting 5d theories are rank-1 theories with the following cubic prepotentials on their Coulomb branches:
\begin{align}
    b=4 \ \ &: \ \ 6\mathcal{F} = 6t_1^2t_2 + 12 t_1 t_2^2 + 8t_2^3 \nonumber \\
    b=5 \ \ &: \ \ 6\mathcal{F} = 6t_1^2t_2 + 6 t_1 t_2^2 + 2t_2^3 \nonumber \\
    b=6 \ \ &: \ \ 6\mathcal{F} = 6t_1^2t_2  \ .
\end{align}

The 5d prepotentials after the Higgsing behave as $\mathcal{F}\sim t_2$ near a boundary $t_2\rightarrow0$ at infinite distance. This means that all these theories have a tower of light KK states coming from the 6d circle in this limit. This is consistent with the fact that they are obtained from 6d theories. However, the behavior of the prepotentials at the other boundary $t_1\rightarrow0$ are different. For the $b=6$ theory, this boundary is located at an infinite distance and the prepotential behaves as $\mathcal{F}\sim t_1^2$, implying the emergence of a critical tensionless string. 

On the other hand, for theories with $b=4,5$,  the boundary at $t_1=0$ lies at a finite distance, and near this point, the potential satisfies the $\Delta_2= 0$ condition. This signals the presence of a rank-1 CFT at the boundary. The specific CFTs can be identified by expressing their prepotentials in terms of K\"ahler parameters dual to the null vectors of the Coulomb branch metrics at the boundary. We find
\begin{align}
    b=4 \ \ &: \ \ 6\mathcal{F} = 8\tilde{t}_1^3 + 8\tilde{t}_2^3 \qquad \rightarrow \quad E_1 \ {\rm CFT}\nonumber \\
    b=5 \ \ &: \ \ 6\mathcal{F} = 2\tilde{t}_1^3 + 2\tilde{t}_2^3 \qquad \rightarrow \quad E_7 \ {\rm CFT}
\end{align}
with $t_1=-2\tilde{t}_1, t_2 = \tilde{t}_2+\tilde{t}_1$ for $b=4$ and $t_1=-\tilde{t}_1, t_2 = \tilde{t}_2+\tilde{t}_1$ for $b=5$. The CS coefficients for the $\tilde{t}_1^3$ terms imply that the local 5d CFTs embedded in the gravity theories are the $E_1$ with a CS coefficient of `8' for $b=4$, and the $E_7$ theory with a CS coefficient of `2' for $b=5$. It should be noted that there are two distinct CFTs, the $E_1$ and $\tilde{E}_1$ CFTs, with a CS coefficient `8', which correspond to the $SU(2)$ gauge theories at theta angle $0$ and $\pi$ respectively. However, the local CFT in the theory with $b=4$ contains only states with even electric charges for the $U(1)$ gauge field of $\tilde{t}_1$, as indicated by the K\"ahler parameter $t_1 = -2\tilde{t}_1$, and thus exhibits a $\mathbb{Z}_2$ 1-form symmetry (though it is broken when the CFT is coupled to gravity). This agrees with the $E_1$ CFT corresponding to $\theta=0$.

These results can be explicitly checked through the genus-0 Gopakumar-Vafa (GV) invariant computations. The genus-0 GV-invariants for the theories with $b\le 8$ were computed in \cite{Hayashi:2023hqa} using the mirror symmetry technique established in \cite{Hosono:1993qy,Hosono:1994ax}. We present a summary of the results for the first few orders in the electric charge expansions in Table \ref{table:P2b4}, Table \ref{table:P2b5}, and Table \ref{table:P2b6} for $b=4, 5$ and $b=6$, respectively. The numbers $(n_0,n_1,n_2)$ in the tables represent the electric charges for the $U(1)$ gauge symmetries in the 5d theories, and the primitive BPS states with charge $(1,0,0), (0,1,0), (0,0,1)$ are dual to the K\"ahler cone generators $t_0, t_1, t_2$, respectively, that appear in the prepotential \eqref{eq:su2-adj}. For more detailed results, readers are referred to \cite{Hayashi:2023hqa}.

\begin{table}[t]
    \scriptsize
    \centering
    \begin{tabular}{c|c|cccc|cccc|cccc}
        & & \multicolumn{12}{c}{$ n_0 $} \\ \hline
        & & $ 0 $ & $ 1 $ & $ 2 $ & $ 3 $ & $ 0 $ & $ 1 $ & $ 2 $ & $ 3 $ & $ 0 $ & $ 1 $ & $ 2 $ & $ 3 $ \\ \hline
        \parbox[t]{1.5ex}{\multirow{7}{*}{\rotatebox[origin=c]{90}{$ n_2 $}}} & $ 0 $ & & $ 4 $ & & & $ 3 $ & $ -8 $ & $ 6 $ & $ -8 $ & $ -6 $ & $ 20 $ & $ -24 $ & $ 12 $  \\
        & $ 1 $ & $ 128 $ & $ 128 $ & & & & $ -256 $ & $ 512 $ & $ 512 $ & & $ 640 $ & $ -2048 $ & $ 2304 $  \\
        & $ 2 $ & $ 4 $ & $ 276 $ & $ 4 $ & & & $ -552 $ & $ 9216 $ & $ 182368 $ & & $ 1380 $ & $ -36872 $ & $ 102312 $ \\
        & $ 3 $ & & $ 128 $ & $ 128 $ & & & $ -256 $ & $ 35328 $ & $ 3120896 $ & & $ 640 $ & $ -141568 $ & $ 1438080 $ \\
        & $ 4 $ & & $ 4 $ & $ 276 $ & $ 4 $ & & $ -8 $ & $ 53246 $ & $ 18166584 $ & & $ 20 $ & $ -213536 $ & $ 7159068 $ \\
        & $ 5 $ & & & $ 128 $ & $ 128 $ & & & $ 35328 $ & $ 47904000 $ & & & $ -141568 $ & $ 17267328 $ \\ \hline
        & & \multicolumn{4}{c|}{$ n_1=0 $} & \multicolumn{4}{c|}{$ n_1=1 $} & \multicolumn{4}{c}{$ n_1=2 $}
    \end{tabular}
    \caption{Genus-0 GV-invariants for $ b=4 $} \label{table:P2b4}
\end{table}

\begin{table}[t]
    \scriptsize
    \centering
    \begin{tabular}{c|c|cccc|cccc|cccc}
        & & \multicolumn{12}{c}{$ n_0 $} \\ \hline
        & & $ 0 $ & $ 1 $ & $ 2 $ & $ 3 $ & $ 0 $ & $ 1 $ & $ 2 $ & $ 3 $ & $ 0 $ & $ 1 $ & $ 2 $ & $ 3 $ \\ \hline
        \parbox[t]{1.5ex}{\multirow{7}{*}{\rotatebox[origin=c]{90}{$ n_2 $}}} & $ 0 $ & & $ 10 $ & & & $ 3 $ & $ -20 $ & $ 45 $ & $ 45 $ & $ -6 $ & $ 50 $ & $ -180 $ & $ 360 $  \\
        & $ 1 $ & $ 140 $ & $ 140 $ & & & & $ -280 $ & $ 1400 $ & $ 23904 $ & & $ 700 $ & $ -5600 $ & $ 18900 $  \\
        & $ 2 $ & $ 10 $ & $ 240 $ & $ 10 $ & & & $ -480 $ & $ 12090 $ & $ 631230 $ & & $ 1200 $ & $  -48380$ & $ 323100 $ \\
        & $ 3 $ & & $ 140 $ & $ 140 $ & & & $ -280 $ & $ 34440 $ & $ 5388840 $ & & $ 700 $ & $ -138040 $ & $ 2336040 $ \\
        & $ 4 $ & & $ 10 $ & $ 240 $ & $ 10 $ & & $ -20 $ & $ 47420 $ & $ 21171435 $ & & $ 50 $ & $ -190160 $ & $ 8162820 $ \\
        & $ 5 $ & & & $ 140 $ & $ 140 $ & & & $ 34440 $ & $ 45601800 $ & & & $ -138040 $ & $ 16345980 $ \\ \hline
        & & \multicolumn{4}{c|}{$ n_1=0 $} & \multicolumn{4}{c|}{$ n_1=1 $} & \multicolumn{4}{c}{$ n_1=2 $}
    \end{tabular}
    \caption{Genus-0 GV-invariants for $ b=5 $} \label{table:P2b5}
\end{table}

\begin{table}[t]
    \scriptsize
    \centering
    \begin{tabular}{c|c|cccc|cccc|cccc}
        & & \multicolumn{12}{c}{$ n_0 $} \\ \hline
        & & $ 0 $ & $ 1 $ & $ 2 $ & $ 3 $ & $ 0 $ & $ 1 $ & $ 2 $ & $ 3 $ & $ 0 $ & $ 1 $ & $ 2 $ & $ 3 $ \\ \hline
        \parbox[t]{1.5ex}{\multirow{7}{*}{\rotatebox[origin=c]{90}{$ n_2 $}}} & $ 0 $ & & $ 18 $ & & & $ 3 $ & $ -36 $ & $ 153 $ & $ 2256 $ & $ -6 $ & $ 90 $ & $ -612 $ & $ 2448 $  \\
        & $ 1 $ & $ 144 $ & $ 144 $ & & & & $ -288 $ & $ 2592 $ & $ 105984 $ & & $ 720 $ & $ -10368 $ & $ 66096 $  \\
        & $ 2 $ & $ 18 $ & $ 216 $ & $ 18 $ & & & $ -432 $ & $ 14112 $ & $ 1356336 $ & & $ 1080 $ & $ -56484 $ & $ 651240 $ \\
        & $ 3 $ & & $ 144 $ & $ 144 $ & & & $ -288 $ & $ 33120 $ & $ 7517856 $ & & $ 720 $ & $ -132768 $ & $ 3145536 $ \\
        & $ 4 $ & & $ 18 $ & $ 216 $ & $ 18 $ & & $ -36 $ & $ 43416 $ & $ 22973544 $ & & $ 90 $ & $ -174096 $ & $ 8712036 $ \\
        & $ 5 $ & & & $ 144 $ & $ 144 $ & & & $ 33120 $ & $ 43396128 $ & & & $ -132768 $ & $ 15493680 $ \\ \hline
        & & \multicolumn{4}{c|}{$ n_1=0 $} & \multicolumn{4}{c|}{$ n_1=1 $} & \multicolumn{4}{c}{$ n_1=2 $}
    \end{tabular}
    \caption{Genus-0 GV-invariants for $ b=6 $} \label{table:P2b6}
\end{table}

The Higgsings can be performed by setting $t_0$, the parameter measuring the $n_0$ charge, to zero. Consequently, the GV-invariants, denoted as $N_{(n_1,n_2)}^{g} = \sum_{n_0=0}^\infty N^g_{(n_0,n_1,n_2)}$, after the Higgsing can be obtained by summing over the GV-invariants with $n_0$ charges for fixed $(n_1,n_2)$ charges. One observes that the GV-invariants in the Higgsed theory at $n_1=0$, which correspond to KK states from 6d massless fields, show an order 2 periodicity: $N_{(0,2n+1)}^{0}=256, N_{(0,2n+2)}^0=284$ for $b=4$; $N_{(0,2n+1)}^0=280, N_{(0,2n+2)}^0=260$ for $b=5$; $N_{(0,2n+1)}^0=288, N_{(0,2n+2)}^0=252$ for $b=6$, with $n\in\mathbb{Z}_{\ge 0}$. This pattern reflects the fact that the Higgsings are taken by giving a vev to a hypermultiplet with $U(1)$ charge $-2$, which results in a $\mathbb{Z}_2$ quotient of the elliptic fibration.

Next, the states with $n_2=0$ in the Higgsed theory become light near the boundary $t_1\rightarrow 0$, and they are in fact the states living in the local CFT at the boundary. For example, when $b=4$, we compute from the table that $N^0_{(1,0)}= \sum_{n_0=0}^\infty N^0_{(n_0,1,0)}=3-8+6-8+3 = -4$ and $N^0_{(2,0)}= -6+20-24+12-8+12-24+20-6 = -4$, and so forth. These GV-invariants $N^0_{(n,0)}$ for $n\in\mathbb{Z}_{>0}$ match precisely with with those for the 5d $E_1$ CFT as expected. Similarly, for $b=5$, the GV-invariants of the Higgsed theory are calculated as $N^0_{(1,0)}=3-20+45+45-20+3 = 56$ and $N^0_{(2,0)}= -6+50-180+360-380+40-380+360-180+50-6 = -272$, and so on. These values agree with the GV-invariants of the 5d $E_7$ CFT. Moving to $b=6$, on the other hand, we compute $N^0_{(1,0)}=2496, N^0_{(2,0)}=223752$, and so on, which show an exponential growth in the number of light particle states as the $n_1$ charge increases. This is consistent with the Hagedorn growth of string oscillator modes in the emergent string limit.

\section{Worldsheet chiral primaries and K\"ahler cone generators}\label{sec:3}

The constraints on the prepotential of 5d supergravity theories, which we discussed in the previous section, can only be applied once the K\"ahler cone of the Coulomb branch moduli space is correctly identified. However, determining the correct K\"ahler cone and its generators is a challenging task. This difficulty stems from the fact that the BPS particle states, which are dual to the K\"ahler cone generators, often arise from non-perturbative objects, such as 5d instantons or 6d strings, that are typically very hard to analyze precisely. For 5d theories derived from circle compactifications of 6d supergravity theories, BPS strings that wrap the Kaluza-Klein circle contribute to 5d BPS states. These states turn out to include primitive BPS states associated with certain K\"ahler cone generators. Therefore, to accurately determine the K\"ahler cone after the 5d reduction, it is crucial to understand the spectrum of BPS strings in the 6d theories. In this section, we will explore spectrum of BPS strings and discuss methods to identify primitive BPS states in circle compactifications of 6d $(1,0)$ supergravity theories.

\subsection{Cone of BPS particle states}

BPS states on the Coulomb branch in a 5d $\mathcal{N}=1$ supergravity can be labeled by their charges under $U(1)$ gauge symmetries at low energy. For those derived from a circle compactification of 6d theories, these $U(1)$ gauge charges can be organized into three groups: $q_I=(q_0, q_\alpha, q_i)$, where $q_0$ is the Kaluza-Klein (KK) charge, $q_\alpha$ represents the tensor charges, and $q_i$ denotes the charges for the 6d gauge symmetries. It is assumed in this context that all non-Abelian gauge symmetries are broken to $U(1)$'s on Coulomb branch of the 5d theory.
The cone of BPS particle states introduced in the previous section is defined as a cone within the electric charge lattice of these $U(1)$ symmetries that contains all BPS particle states and features at least one BPS state situated on each boundary of the cone. Among the states on these boundaries, those whose electric charge cannot be expressed as multiples of the charges of other BPS states are referred to as {\it primitive BPS states}. These primitive BPS states are generators of the cone of BPS particle states.

After performing a circle compactification of a 6d supergravity theory, we can immediately identify certain primitive BPS states in the resulting 5d theory from the massless spectrum of the original 6d theory. The 5d theory has BPS states from 6d hyper and vector multiplets that do not carry KK charges, say $q_0=0$. Specifically, among these, the states with the minimal (positive) gauge charge (or with a minimal dominant weight for a non-Abelian gauge group) in the sublattice of the weight lattice occupied by 6d massless charged fields are primitive BPS states. Also, when $q_0>0$, BPS states from the 6d massless fields can have negative gauge charges. Among these, the states with the most negative gauge charges and a unit KK charge are identified as primitive BPS states. These primitive states serve as generators for the cone of BPS particle states in the 5d theory. All other BPS states from the 6d massless spectrum with or without carrying KK charges, and that share the same supercharges can be obtained by positive combinations of these primitive BPS states. 


Let us illustrate this with a simple example. Consider an $SU(2)$ gauge group coupled to a number of fundamental hypermultiplets. In this case, we can identify two primitive BPS states with $U(1)\subset SU(2)$ gauge charges derived from the 6d perturbative massless spectrum. At a generic point on the Coulomb branch of the 5d theory, we can choose a supercharge such that the BPS states with respect to the supercharge have electric charges $(q_0,q_{U(1)})=(n,1)$ or $(n+1,-1)$ when they originate from the fundamental hypermultiplets, or $(q_0,q_{U(1)})=(n,2)$ or $(n+1,-2)$ when they are from the vector multiplet, where $n\in\mathbb{Z}_{\ge0}$. Therefore, the cone of these BPS states is generated by two primitive states with $(q_0,q_{U(1)})=(0,1)$ and $(1,-2)$.

Additionally, the spectrum of the 5d theory also includes BPS states that carry 6d tensor charges. These states arise from excitations of BPS strings wrapped around the 6d circle. The spectra of such BPS string excitations in local 6d theories, including 6d superconformal field theories and little string theories, have been extensively studied recently. For more details, we refer to \cite{Haghighat:2014vxa,DelZotto:2018tcj,Kim:2020hhh,Kim:2023glm} and the references therein. However, it is important to note that such strings and their spectra in local 6d theories do not contain the entire range of BPS strings in gravity theories. 

In 6d supergravity theories, there exists a distinct class of BPS strings, often referred to as {\it supergravity strings}, as introduced in \cite{Kim:2019vuc,Katz:2020ewz}. These strings are associated with 5d BPS black holes under a circle compactification, and also Nef divisors in M-theory compactifications on a compact CY3. The spectrum of these strings in a specific subset of 6d theories that admit geometric realizations in F-theory can be computed using the mirror symmetry technique, as introduced in \cite{Hosono:1993qy,Hosono:1994ax}.  However, this method is not applicable to all supergravity theories. This leads to a significant need for alternative approaches to explore the spectrum of strings in more generic 6d theories and to accurately identify the correct cone of 5d BPS states. Developing such methods is crucial for applying constraints on 6d supergravity theories in our study.

\subsection{BPS strings and ground state energy}

BPS strings have played crucial roles in understanding non-perturbative aspects of 6d and 5d theories. In particular, supergravity strings have been used in the Swampland program for constraining low-energy effective theories in various dimensions, as demonstrated in studies such as \cite{Kim:2019vuc,Lee:2019skh,Kim:2019ths,Katz:2020ewz,Tarazi:2021duw,Martucci:2022krl,Hamada:2024oap}. In 6d theories, supergravity strings are specifically defined as BPS strings with a charge $Q$ that meets the conditions \cite{Kim:2019vuc}
\begin{align}
    Q\cdot J > 0 \ , \quad c_L, c_R\ge0 \ , \quad k_l, k_i \ge 0 \ , 
\end{align}
where $c_{L,R}$ are the left and right central charges of the 2d worldsheet CFT on the string, taking the values of
\begin{align}
    c_L = 3Q\cdot Q-9Q\cdot a + 2 \ , \quad c_R = 3Q\cdot Q - 3Q\cdot a \ ,
\end{align}
and $k_l, k_i$ are the levels of the current algebras for the $SU(2)_l$ flavor symmetry and the 6d bulk gauge symmetry $G_i$, given by
\begin{align}
    k_l = \frac{1}{2}(Q\cdot Q + Q\cdot a + 2) \ , \quad k_i = Q\cdot b_i \ .
\end{align}
We note that the worldsheet theory on the supergravity theory is a $\mathcal{N}=(0,4)$ SCFT with an $SU(2)_R$ R-symmetry and an $SU(2)_l$ flavor symmetry, which come from the $SO(4)$ Lorentz symmetry along transverse $\mathbb{R}^4$ directions. In the central charge formulae above, we removed the contributions from the center of mass degrees of freedom, which are
\begin{align}
    c_L^{\rm c.o.m} = 4 \ ,\quad c_L^{\rm c.o.m} = 6 \ , \quad k_l^{\rm c.o.m} = -1 \ ,\quad k_i^{\rm c.o.m} = 0  \ .
\end{align}

The 5d BPS particles from such a string are captured by the elliptic genus of the 2d $(0,4)$ worldsheet CFT wrapped on the 6d circle. The elliptic genus of a supergravity string with tensor charge $Q$ is defined as a weighted trace over the BPS spectrum in the Ramond (R) sector, and it can be written as
\begin{align}\label{eq:elliptic-genus}
    Z_Q(q,y,z_i) = {\rm Tr}_{R}(-1)^F q^{L_0} \bar{q}^{\bar{L}_0} y^{J_l}z_i^{f_i} \ ,
\end{align}
where $F$ denotes the fermion number, $L_0, \bar{L}_0$ are the left-moving and right-moving conformal dimensions respectively, $J_l$ is the Cartan generator of the $SU(2)_l$ symmetry, and $f_i$ denotes the Cartan generators of the 6d bulk gauge symmetries $G_i$. Due to the index nature of this elliptic genus, only ground states in the right-moving sector with $\bar{L}_0=0$ contribute, and thus the elliptic genus is independent of $\bar{q}$. The KK charges of the 5d BPS states are identified with the conformal dimensions $L_0$ of worldsheet states, i.e. $q_0 = L_0$.

The spectra of primitive BPS strings are particularly interesting because their ground states provide primitive BPS states in the 5d reduction that carry tensor charges. Here, primitive BPS strings are defined as those whose tensor charges cannot be obtained from a positive combination of tensor charges of other BPS strings. 
Naively, one might expect that the ground state of a primitive string with charge $Q_\alpha$ would yield a 5d BPS particle state with charge $(q_0,q_\alpha,q_i) = (0,Q_\alpha,0)$. However, this naive expectation turns out to be wrong because vacuum Casimir energy induces non-zero KK charge for each ground state of 6d BPS strings, which leads to non-zero $q_0$ for the corresponding 5d BPS state. Thus, it is crucial to carefully consider the KK charge shift when examining the string excitations.

The KK charges of the ground states of wrapped BPS strings can be systematically calculated using certain properties of 2d CFTs and of BPS strings in 6d supergravity. First, we note that KK charges of 5d BPS states from the 6d string spectrum are identified with $L_0$ eigenvalues of the states in the R sector. Also, we know that the vacuum state in Neveu-Schwarz (NS) sector has an eigenvalue $L_0 = -\frac{c_L}{24}$. Then, $L_0$ eigenvalues of the states in the R sector can be obtained by a spectral flow \cite{Lerche:1989uy}. Furthermore, we note that the fermion number operator of the elliptic genus for the supergravity strings can be realized by $F=J_l+J_R$ where $J_l,J_R$ are two Cartan generators of the $SU(2)_l\times SU(2)_R \cong SO(4)$ transverse rotations. With these factors together, we find that the spectral flow can be effectively implemented as
\begin{align}
    L_0 \ \rightarrow \ L_0 + \frac{k_l}{4} \ , \quad \bar{L}_0 \ \rightarrow \ \bar{L}_0 + \frac{k_R}{4} \ ,
\end{align} 
where $k_R = c_R/6$ is the level for the current algebra of $SU(2)_R$ R-symmetry. Then, the ground state in R sector has vacuum energies as
\begin{align}
    L_0 &= -\frac{c_L}{24}+\frac{k_l}{4} = \frac{Q\cdot a}{2}+\frac{1}{6} \ , \nonumber \\
    \bar{L}_0 &= -\frac{c_R}{24} + \frac{k_R}{4} = 0 \ .
\end{align}

Therefore, the ground state of a supergravity string with charge $Q_\alpha$ has electric charges 
\begin{align}\label{eq:string-BPS-particle}
    (q_0,q_\alpha,q_i) = (\frac{Q\cdot a}{2},Q_\alpha,0) \ ,
\end{align}
under the 5d reduction. Here, we included the center of mass contribution to the vacuum energy $L_0^{\rm c.o.m}=-1/6$. The KK charge shift $q_0=Q\cdot a/2$ for the ground state of a string with charge $Q$ agrees with that of BPS states, which corresponds to the D0 charge induced on the D4 brane, in the 4d supergravity theory engineered in Type IIA compactification as studied in \cite{Vafa:1995zh,Green:1996dd}. We claim that each primitive BPS string with tensor charge $Q_\alpha$ provides a primitive BPS particle state with electric charge \eqref{eq:string-BPS-particle} in the 5d theory.

As an example, let us discuss the cone of BPS particle states in the $T=0$ theory with $SU(2)$ gauge group and matter content given in \eqref{eq:su2-hyper}, compactified on a circle, which we discussed in the previous section.  After reducing to 5d, we have BPS particle states with electric charges $(q_0,q_1,q_2)$ where $q_0$ is the KK charge, $q_1$ is the tensor charge, and $q_2$ is the charge under the $U(1)\subset SU(2)$ gauge symmetry. The massless spectrum of the 6d theory includes charged fields which have electric charges $\pm1$ and $\pm2$ for $U(1)\subset SU(2)$ gauge group. These massless fields provide two primitive BPS states in 5d with charges $(0,0,1)$ and $(1,0,-2)$, following the earlier discussions. Additionally, the ground state of a BPS string with tensor charge $Q=1$ gives rise to another primitive BPS state with charge $(-\frac{3}{2},1,0)$. These three primitive BPS states serve as generators of the cone of BPS particle states, although there may be additional generators as we will discuss in the next subsection. These states indeed correspond to the primitive BPS states with charges $(n_0,n_1,n_2)=(0,0,1), (1,0,0)$, and $(0,1,0)$ respectively in the GV-invariants listed in Table \ref{table:P2b4}, \ref{table:P2b5}, \ref{table:P2b6}.

The K\"ahler cone is formed by real scalar parameters that are dual to the charges of these primitive states. Thus, we identify three K\"ahler cone generators in this theory given by
\begin{align}\label{eq:three-generators}
    t_0 = \phi_0-2\phi_2 \ , \quad t_1 = \phi_1 - \frac{3}{2}\phi_0 \ ,\quad t_2 = \phi_2 \ ,
\end{align}
where $\phi_i$ denotes the scalar coordinate (or scalar vev) along charge $q_i$ direction in the Coulomb branch. We can compute the prepotential of the 5d theory in terms of these K\"ahler cone generators, $t_0,t_1,t_2$. From the effective CS Lagrangian in \eqref{eq:Lagrangian} with $\Omega=1$ and $a=-3$, we compute
\begin{align}\label{eq:pre-su2-gravity}
    6\mathcal{F}_{SU(2)_b} =&\ \frac{9}{4}\phi_0^3  + 3\phi_0 \phi_1^2-6b(\phi_1 +\frac{3}{2}\phi_0)\phi_2^2 - (8-n_{\bf2} - 8n_{\bf 3})\phi_2^3 \nonumber\\
    =& \ 9 t_0^3 + 9t_0^2 t_1 + 3 t_0t_1^2 +54 t_0^2 t_2 + 6 t_1^2 t_2 +36t_0t_1t_2 \nonumber \\
    & +18(6-b)t_0t_2^2 + 6(6-b)t_1t_2^2 + 2(6-b)^2 t_2^3 \ .
\end{align}
This result, for $1\le b\le 8$, perfectly matches with the prepotential presented in \eqref{eq:su2-adj}, which was derived from the triple intersection numbers of Nef divisors with K\"ahler parameters $t_{0,1,2}$ in the associated Calabi-Yau 3-folds. 

The CY3 geometries for $1\le b\le 6$ have simplicial K\"ahler cones generated by $t_0,t_1,t_2\ge0$ that appear in the prepotential. Indeed, the cubic CS coefficients of these K\"ahler parameters are non-negative. This is expected for the K\"ahler parameters for Nef divisors in geometry. However, as discussed in \cite{Hayashi:2023hqa}, the K\"ahler cones of the CY3 geometries for $b=7,8$ are non-simplicial. One may notice that the cubic CS coefficients of terms like $t_0t_2^2$ and $t_1t_2^2$ are negative in these geometries. This does not mean that these geometries are inconsistent or that the K\"ahler cone generators $t_0,t_1,t_2$ are misidentified. In fact, these geometries have an additional K\"ahler cone generator, and thus the actual K\"ahler cone is smaller than one generated by just three generators $t_0,t_1,t_2$. It turns out that these extra K\"ahler cone generators arise from excitations of BPS strings in the 6d supergravity theory. We will now discuss these extra K\"ahler cone generators from 6d BPS strings. 

\subsection{Chiral primaries and extra K\"ahler cone generators}\label{sec:chiralprimary}

The ground state of each primitive BPS string, as mentioned above, gives rise to a primitive BPS state in the 5d reduction of a 6d supergravity theory. This raises a natural question: Can other oscillator modes of BPS strings also contribute additional primitive BPS states in 5d, and consequently, more K\"ahler cone generators than those we identified in the previous subsections? If this is the case, the K\"ahler cone for the moduli space of the 5d theory would be smaller than we naively expect, and thus we need to be more cautious when applying positivity constraints on the prepotential of the low energy effective theory.

BPS strings in 6d supergravity theories are non-perturbative objects, and there is currently no standard method to fully determine the spectra of their worldsheet CFTs. As such, precisely identifying the exact spectrum of BPS states arising from strings wrapped around the 6d circle remains a highly challenging task. Therefore, our strategy will be to focus on identifying {\it candidate primitive states}, which will be defined shortly, from the worldsheet excitations, rather than attempting to determine the exact BPS spectrum from the strings. This more manageable task can be accomplished by utilizing the unitarity of 2d CFTs that host flavor current algebras associated with 6d bulk gauge group. We will now illustrate this approach using 6d theories with $U(1)$ or $SU(2)$ gauge symmetry.

First, consider a $T=0$ theory with an $SU(2)$ gauge symmetry. The 2d worldsheet CFT on a BPS string with tensor charge $Q$ supports an $SU(2)_k$ current algebra at level $k\equiv Qb$, where $b$ is the anomaly coefficient for the $SU(2)$ gauge algebra appearing in the 6d anomaly cancellation. The current algebra is realized in the left-moving sector and is generated by chiral primary fields $j^3,j^\pm$ with conformal weight $h=1$. The spectrum of WZW model associated with the $SU(2)$ group at level $k$ consists of $k+1$ primary fields corresponding to the irreducible representations of $SU(2)$. These primaries are labeled by spin $j/2$, where $0\le j\le k$, and their conformal weights are \cite{DiFrancesco:1997nk}
\begin{align}
    h_j = \frac{j(j+2)}{4(k+2)} \ .
\end{align}

The primary fields in the WZW model can provide BPS states captured in the elliptic genus in \eqref{eq:elliptic-genus}. For this, the primary field should be combined with other singlet fields, which have positive conformal weights by unitarity, to form a BPS state with an integral conformal dimension. Thus, a BPS state arising from a WZW primary field of spin $j/2$ should exhibit a bigger conformal dimension than $h_j$. This implies that when the 6d theory is reduced to 5d, the WZW primary field of spin $j/2$ in the 2d worldsheet CFT for a tensor charge $Q$ can provide 5d BPS particle states with electric charges
\begin{align}
    (q_0,q_1,q_2) = (Q\cdot a/2,Q,0) + (n,0,\ell) \ ,
\end{align}
where $n\in\mathbb{Z}_{\ge0}$ denotes the left-moving excitation level satisfying $n\ge h_j$, and $-j\le\ell\le j$ for $0\le j\le k$. Here, the ground state corresponds to $n=\ell=0$.

Currently, there is no known method to determine the exact value of $n$ for generic primary states in worldsheet CFTs. However, if $n$ is sufficiently small, some of these states, in particular those with $\ell=-j$, can become extra generators of the cone of BPS particle states in 5d. On the other hand, for large values of $n$, specifically when $n\ge j/q_{\rm max}$, the electric charge of such a state can generally be expressed by a positive combination of the charges of the primitive BPS states that we have already identified from the ground states of BPS strings and the 6d massless spectrum. Here, $q_{\rm max}$ denotes the maximal Cartan charge of 6d $SU(2)$ massless fields. In fact, the smaller the value of $n$, the higher the potential for these states being extra primitive BPS states. Bearing this in mind, we now define {\it candidate primitive states} as the BPS states that could potentially arise from the spectrum of BPS strings in 6d and become primitive BPS states in 5d with specific electric charges 
\begin{align}\label{eq:candidate-BPS}
    (q_0,q_1,q_2) = (Q\cdot a/2,Q,0) + (m,0,-j) \quad {\rm with}\quad h_j\le m < \frac{j}{q_{\rm max}}
\end{align}
with $m\in\mathbb{Z}_{>0}$. The charges of these states, when exist, cannot be expressed as positive combinations of the charges of the other primitive states, and thus become additional primitive states.


Similarly, when the 6d gauge group is $U(1)$, the worldsheet CFT with a $U(1)$ current algebra at level $k$ may contain the primary fields with $U(1)$ charge $\ell$ and conformal weight \cite{DiFrancesco:1997nk}
\begin{align}
    h_\ell = \frac{\ell^2}{2k} \ .
\end{align}
Here, the $U(1)$ charge, denoted as $\ell$, is integral and satisfies $-k/2\le \ell \le k/2$, which is determined by applying a quantization condition and a charge conjugation symmetry that we expect for the worldsheet CFTs on BPS strings in 6d. These worldsheet primary fields can lead to primitive BPS states with the charge given in \eqref{eq:candidate-BPS} by substituting $h_j$ and $j$ with $h_\ell$ and $\ell$, respectively.

Before we discuss the implications of the candidate primitive states on the Coulomb branch analysis with some examples, let us make a few comments related to them. Firstly, descendants of a primary field cannot yield new primitive BPS states. This is because the corresponding states are generated by acting $j^3, j^\pm$ operators on the primary state and, since this action increases conformal weight by $+1$ while changing $SU(2)$ charge by at most $\pm2$, the charge of such a state can always be expressed by positive combinations of the charge of the primary state and those of other primitive BPS state from the 6d massless multiplets in the models under consideration. This also holds true for the $U(1)$ gauge symmetry as well, as evidenced by the characters of the $U(1)$ current algebra.

Second, there might be concerns about other BPS states derived from primary fields dressed by bosonic zero modes on the instanton moduli space that can arise when $Q\ge b$. For a $U(1)$ gauge group, since the bosonic zero modes of $U(1)$ instantons do not carry gauge charges, such states cannot contribute additional primitive BPS states. For an $SU(2)$ gauge group, however, we cannot ignore the possibility of additional primitive BPS states from such states. Thus, in this paper, we will focus on the $SU(2)$ gauge theories that contain adjoint hypermultiplets and thus can be Higgsed to $U(1)$ gauge theories. In these cases, the Higgsing process ensures that worldsheet primary fields dressed by bosonic zero modes of instantons before the Higgsing cannot yield additional primitive BPS states.

Let us now present some examples of constructions in string theory and discuss the role of the candidate primitive states discussed above. Consider the $T=0$ theory with an $SU(2)$ gauge symmetry with matters given in \eqref{eq:su2-hyper}. As previously mentioned, when $b=7,8$, these theories have non-simplicial K\"ahler cones in their 5d reductions. These theories have toric hypersurface constructions, and their GV-invariants can be computed using the mirror symmetry technique, as carried out in \cite{Hayashi:2023hqa}. The genus-0 GV-invariants for these theories are summarized in Table \ref{table:P2b7-gv0} and \ref{table:P2b8-gv0}. 

From the results of these GV invariants, we can see that the cone of BPS states is generated by states with unit charges $(n_0,n_1,n_2)=(1,0,0), (0,1,0)$, and $(0,0,1)$, which correspond respectively to the KK state from the 6d vector (or adjoint hyper) multiplet, the ground state of the string with tensor charge $Q=1$, and the 6d fundamental hypermultiplets. The prepotential in \eqref{eq:su2-adj} is written in terms of the K\"aher parameters $t_{0,1,2}$ that are dual to these states.  

\begin{table}[t]
    \scriptsize
    \centering
    \begin{tabular}{c|c|cccc|cccc|cccc}
        & & \multicolumn{12}{c}{$ n_0 $} \\ \hline
        & & $ 0 $ & $ 1 $ & $ 2 $ & $ 3 $ & $ 0 $ & $ 1 $ & $ 2 $ & $ 3 $ & $ 0 $ & $ 1 $ & $ 2 $ & $ 3 $ \\ \hline
        \parbox[t]{1.5ex}{\multirow{7}{*}{\rotatebox[origin=c]{90}{$ n_2 $}}} & $ -1 $ & & & & & & & & $ 56 $ &  \\
        & $ 0 $ & & $ 28 $ & & & $ 3 $ & $ -56 $ & $ 378 $ & $ 12747 $ & $ -6 $ & $ 140 $ & $ -1512 $ & $ 9828 $  \\
        & $ 1 $ & $ 140 $ & $ 140 $ & & & & $ -280 $ & $ 3920 $ & $ 286048 $ & & $ 700 $ & $ -15680 $ & $ 158760 $  \\
        & $ 2 $ & $ 28 $ & $ 204 $ & $ 28 $ & & & $ -408 $ & $ 15330 $ & $ 2253314 $ & & $ 1020 $ & $ -61376 $ & $ 1039248 $ \\
        & $ 3 $ & & $ 140 $ & $ 140 $ & & & $ -280 $ & $ 31920 $ & $ 9341976 $ & & $ 700 $ & $ -127960 $ & $ 3806460 $ \\
        & $ 4 $ & & $ 28 $ & $ 204 $ & $ 28 $ & & $ -56 $ & $ 40274 $ & $ 24037573 $ & & $ 140 $ & $ -161504 $ & $ 9000684 $ \\
        & $ 5 $ & & & $ 140 $ & $ 140 $ & & & $ 31920 $ & $ 41386464 $ & & & $ -127960 $ & $ 14735700 $ \\ 
        & $ 6 $ & & & $ 28 $ & $ 204 $ & & & $ 15330 $ & $ 49434828 $ & & & $ -61376 $ & $ 17309160 $ \\
        & $ 7 $ & & & & $ 140 $ & & & $ 3920 $ & $ 41386464 $ & & & $ -15680 $ & $ 14735700 $ \\ \hline
        & & \multicolumn{4}{c|}{$ n_1=0 $} & \multicolumn{4}{c|}{$ n_1=1 $} & \multicolumn{4}{c}{$ n_1=2 $}
    \end{tabular}
    \caption{Genus-0 GV-invariants for $ b=7 $} \label{table:P2b7-gv0}
\end{table}

\begin{table}[t]
    \scriptsize
    \centering
    \begin{tabular}{c|c|cccc|cccc|cccc}
        & & \multicolumn{12}{c}{$ n_0 $} \\ \hline
        & & $ 0 $ & $ 1 $ & $ 2 $ & $ 3 $ & $ 0 $ & $ 1 $ & $ 2 $ & $ 3 $ & $ 0 $ & $ 1 $ & $ 2 $ & $ 3 $ \\ \hline
        \parbox[t]{1.5ex}{\multirow{7}{*}{\rotatebox[origin=c]{90}{$ n_2 $}}} & $ -2 $ & & & & & & & & $ -4 $ &  \\
        & $ -1 $ & & & & & & & & $ 256 $ &  \\
        & $ 0 $ & & $ 40 $ & & & $ 3 $ & $ -80 $ & $ 780 $ & $ 46224 $ & $ -6 $ & $ 200 $ & $ -3120 $ & $ 29640 $  \\
        & $ 1 $ & $ 128 $ & $ 128 $ & & & & $ -256 $ & $ 5120 $ & $ 572672 $ & & $ 640 $ & $ -20480 $ & $ 299520 $  \\
        & $ 2 $ & $ 40 $ & $ 204 $ & $ 40 $ & & & $ -408 $ & $ 16128 $ & $ 3211024 $ & & $ 1020 $ & $ -64592 $ & $ 1433520 $ \\
        & $ 3 $ & & $ 128 $ & $ 128 $ & & & $ -256 $ & $ 30720 $ & $ 10856448 $ & & $ 640 $ & $ -123136 $ & $ 4334208 $ \\
        & $ 4 $ & & $ 40 $ & $ 204 $ & $ 40 $ & & $ -80 $ & $ 37874 $ & $ 24626000 $ & & $ 200 $ & $ -151904 $ & $ 9127368 $ \\
        & $ 5 $ & & & $ 128 $ & $ 128 $ & & & $ 30720 $ & $ 39596032 $ & & & $ -123136 $ & $ 14073984 $ \\ 
        & $ 6 $ & & & $ 40 $ & $ 204 $ & & & $ 16128 $ & $ 46253880 $ & & & $ -64592 $ & $ 16214040 $ \\
        & $ 7 $ & & & & $ 128 $ & & & $ 5120 $ & $ 39596032 $ & & & $ -20480 $ & $ 14073984 $ \\ \hline
        & & \multicolumn{4}{c|}{$ n_1=0 $} & \multicolumn{4}{c|}{$ n_1=1 $} & \multicolumn{4}{c}{$ n_1=2 $}
    \end{tabular}
    \caption{Genus-0 GV-invariants for $ b=8 $} \label{table:P2b8-gv0}
\end{table}

However, there are fourth primitive states with charges $(n_0,n_1,n_2)=(3,1,-1)$ for $b=7$ and $(3,1,-2)$ for $b=8$. Since these states carry negative $n_2$ charges, they cannot be expressed as positive combinations of other primitive states. These states come from the worldsheet primary fields on a string with charge $Q=1$, as indicated by $n_1=1$. More specifically, the primary fields corresponding to these states have a conformal weight $h=3$, which is because $n_0=3$ for these states, and spins $j=7$ and $j=8$ respectively. Here, the spins are determined by their $U(1)\subset SU(2)$ charge $-7$ and $-8$, which can be deduced using the charge conjugation symmetry (or the reflection symmetry along $n_2$ axis) of the GV-invariants in Table \ref{table:P2b7-gv0} and \ref{table:P2b8-gv0}. Indeed, these states fall within the range of candidate primitive states with charge $m=3$ and $j=7,8$ given in \eqref{eq:candidate-BPS}, which satisfy the criteria for primitive states as
\begin{align}
    b=7 \ \ : \ \  7/4\le 3 < 7/2 \ \ , \qquad \quad b=8 \ \ : \ \  8/4\le 3 < 8/2 \ ,
\end{align}
with $h_7=7/4$ for $b=7$ and $h_8=8/4$ for $b=8$, where $q_{\rm max}=2$, the maximal gauge charge of the adjoint hypermultiplet.

It then follows that the K\"ahler parameter dual to the extra primitive state gives rise to the fourth generator of the K\"ahler cone, defined respectively as
\begin{align}
    b=7 \ \ : \ \ t_4 = t_1+3t_0-t_2 \ \ , \qquad \quad b=8 \ \ : \ \ t_4 = t_1+3t_0-2t_2 \ .
\end{align}
Indeed, these generators are not independent of the other generators $t_{0,1,2}$ defined in \eqref{eq:three-generators}. In particular, one may notice that somewhere in the cone with $t_{0,1,2}\ge0$, the fourth generator can become negative, implying that the actual K\"ahler cone $\mathcal{K}$ is smaller than the cone $\mathcal{C}(t_i)$ defined by $t_{0,1,2}\ge0$, i.e. $\mathcal{K} \subset \mathcal{C}(t_{i})$. Thus,  the presence of negative CS coefficients in the prepotential \eqref{eq:su2-adj} written in terms of $t_{0,1,2}\ge0$ does not conflict with the positivity constraints on CS coefficients, as presented in \eqref{eq:CScoeff-positivity}, within the Coulomb branch. 

One can check that the cubic CS coefficients always satisfy the positivity constraints within the K\"ahler cone formed by four generators $t_{0,1,2}\ge0$ and $t_4\ge0$. For example, we can rewrite the prepotential in terms of $t_0,t_2$ and $\tilde{t}_1=t_1-2t_2$ instead as follows:
\begin{align}
    6\mathcal{F}_{SU(2)_b} =&\ 9 t_0^3 + 9t_0^2 \tilde{t}_1 + 3 t_0\tilde{t}_1^2 +72 t_0^2 t_2 + 6 \tilde{t}_1^2 t_2 +48t_0\tilde{t}_1t_2 \nonumber \\
    & +6(32-3b)t_0t_2^2 + 6(10-b)\tilde{t}_1t_2^2 + 2(b^2-18b+84) t_2^3 \ ,
\end{align}
with $b=7,8$. Now, the cone generated by positive $t_0,\tilde{t}_1,t_2$ is contained within the actual K\"ahler cone, i.e. $\mathcal{C}(t_0,\tilde{t}_1,t_2)\subset \mathcal{K}$. Thus as expected, all CS coefficients in this expression are positive.

\section{Swampland for $U(1)$ and $SU(2)$ models}\label{sec:4}
In this section, we will examine the Coulomb branch moduli spaces of 6d supergravity theories characterized by $T=0$ and a single rank-1 gauge factor compactified on a circle. This analysis will utilize the consistency conditions for effective prepotentials in 5d that we introduced earlier in Section \ref{sec:2}. By examining various examples, we will show that the conditions for anomaly cancellations are robust, such that most 6d anomaly-free theories we have examined may potentially have consistent cubic prepotentials in their 5d reductions, and thus our analysis cannot definitively exclude them. However, we will also highlight that certain classes of anomaly-free theories that lack the expected 1-form gauge symmetry from their massless spectra violate the constraints on their K\"ahler moduli space, placing them in the Swampland. It is important to note that in this section, we will use only the low-energy aspects of supergravity theory without assuming any particular UV completions, such as string theory or F-theory compactifications.

\subsection{Infinite families of Abelian theories}

It is known that the number of consistent 6d supergravity theories that can be engineered in F-theory is finite. Also, the number of anomaly-free massless spectra for 6d supergravity theories with only non-Abelian gauge groups is proven to be finite when $T<9$ \cite{Kumar:2009ae,Kumar:2010ru}. This raises the question of whether the finiteness of consistent massless spectra extends to more general cases, such as those including Abelian gauge groups or additional tensor fields with $T\ge9$. However, there are already known exceptions to this limitation. Specifically, certain infinite families of anomaly-free massless spectra were identified even at $T=0$ in \cite{Taylor:2018khc}. They have a single $U(1)$ gauge factor and a number of matter fields whose charges are tuned to cancel all anomalies.  It remains uncertain whether these theories can be consistently coupled to gravity beyond the anomaly cancellation consideration, and thus can be classified as part of the landscape.\footnote{Additional constraints on this infinite family of $U(1)$ theories, derived from the Dai-Freed anomaly cancellation, were also discussed in \cite{Basile:2023zng}.} In this subsection, we will investigate the Coulomb branches of these infinite families and show that some of them fall into the Swampland.

An infinite family of anomaly-free models studied in \cite{Taylor:2018khc} includes the $T=0$ theories with a $U(1)$ gauge group that contain 54 hypermultiplets with $U(1)$ charges $p,r,$ and $p+r$, and can be summarized as
\begin{align}
    54\times (\pm p)+54\times (\pm r) + 54\times(\pm(p+r)) \ , \quad a= -3\ , \quad b = 6(p^2+pr+r^2) \ ,
\end{align}
with $p,r\in\mathbb{Z}$.

To examine these theories, let us first compactify them to five-dimensions and move along the Coulomb branch in the resulting 5d theory. The 5d theory has three $U(1)$ gauge fields corresponding to the gravi-photon, the 6d $U(1)$ gauge field, and the gauge field dual to the 6d self-dual 2-form tensor field. Based on the discussions in the previous section, we can easily identify three primitive BPS states: one is from the 6d hypermultiplet with the minimum $U(1)$ charge, another is from a KK state with largest negative $U(1)$ charge, and the last one is from the ground state of the string with tensor charge $Q=1$. Their electric charges can be expressed as
\begin{align}\label{eq:U1-generators}
    (q_0,q_1,q_2) = (0,0,p) \ , \quad (1,0,-p-r) \ , \quad (-3/2,1,0) \ ,
\end{align}
respectively, with $0<p\le r$. Here, $q_0, q_1, q_2$ denote KK charge, tensor charge, and the 6d $U(1)$ gauge charge respectively. For the last charge, the KK charge shift, which is $Q\cdot a/2 = -3/2$, for the ground state of the string with unit tensor charge $Q=1$ is taken into account.
The cubic prepotential on the Coulomb branch of the 5d theory can be computed using the formula given in \eqref{eq:Lagrangian}. We compute it as
\begin{align}
    6\mathcal{F} = \frac{9}{4}\phi_0^3 + 3\phi_0 \phi_1^2 -3b(\phi_1+\tfrac{3}{2}\phi_0)\phi_2^2 + 27(p^3+r^3+(p+r)^3)\phi_2^3 \ ,
\end{align}
where we used the K\"ahler parameters $\phi_i$ in a basis such that the masses of 5d BPS states are given by $M \sim \sum_{i=0}^2 q_i \phi_i$. 

The constraints on a cubic prepotential can be easily checked when the prepotential is expressed in terms of the parameters that generate the K\"ahler cone. These generators of the K\"ahler cone are dual to the primitive states we have just identified, which are
\begin{align}\label{eq:u1-t012}
    t_0=\phi_0-(p+r)\phi_2 \ , \quad t_1=\phi_1 - 3/2\phi_0 \ , \quad t_2 = \phi_2 \ .
\end{align}
Using these parameters, we can then reformulate the prepotential as
\begin{align}
    6\mathcal{F} =&\ 9t_0^3+9t_0^2 t_2+3t_0t_1^2 + 3(p+r)\left(9t_0^2+t_1^2+6t_0t_1\right)t_2  \nonumber \\
    &- 9(p^2+r^2)\left(3t_0+t_1\right)t_2^2+9(p^3+r^3)t_2^3 \ .
\end{align}
We observe that this prepotential includes terms with negative coefficients. This indicates that the K\"ahler cone is non-simplicial, or the theory may otherwise be inconsistent. Indeed, higher level excitations of BPS strings might yield additional BPS states that cannot be expressed as positive combinations of the primitive states we identified earlier in \eqref{eq:U1-generators}. We will discuss this issue shortly.

Now, we will Higgs this theory by giving a vev to the hypermultiplet with charge $(1,0,-p)$, which is one of KK states coming from the 6d hypermultiplet with charge $\pm p$. We can implement this Higgsing by taking the limit $\phi_0-p\phi_2=t_0+rt_2\rightarrow 0$ at the level of the K\"ahler parameters. Thus, setting $t_0=-rt_2$ in the prepotential gives rise to the prepotential of the Higgsed theory. 

However, when we take this limit, we need to be careful about the presence of hypermultiplets whose masses approach zero. For these hypermultiplets, we must perform flop transitions which lead us to another chamber where the BPS and anti-BPS states from the hypermultiplets are swapped. The flop transition for a hypermultiplet with mass proportional to $q_it_i>0$ in one chamber to another chamber where $q_it_i<0$ can be realized in the prepotential as
\begin{align}
     \mathcal{F}_{q_it^i<0}= \mathcal{F}_{q_it^i>0} - \frac{1}{6}(q_it^i)^3 \ .
\end{align}

The number of hypermultiplets that need to be flopped depends on the relative ranges of $p$ and $r$. Let us first assume that $r\ge q>r/2$. Then three KK hypermultiplet states with charges $(1,0,-p-r), (1,0,-r)$, and $(2,0,-p-r)$ respectively, will require flop transitions as we approach the limit $t_0\rightarrow -r t_2$. Considering the flop transitions for these hypermultiplets, we then calculate the prepotential after the Higgsing as follows:
\begin{align}
    6\mathcal{F} = 3p t_1^2t_2-9(p^2+2pr+2r^2)t_1t_2^2+9(24r^3-33pr^2+39p^2r-11p^3)t_2^3 \ .
\end{align}
The negative CS coefficient of the $t_1t_2^2$ term here may imply that the K\"ahler cone is smaller than the cone generated by positive values of $t_1$ and $t_2$. Let us investigate this possibility from now on.

Additional primitive BPS states might arise from chiral primary fields in the 2d CFTs on 6d BPS strings. For the $U(1)$ models, the worldsheet CFT for a tensor charge $Q$ contains a $U(1)$ current algebra at level $k=Qb$. As discussed around \eqref{eq:candidate-BPS}, the worldsheet primary fields could yield candidate primitive states with electric charges
\begin{align}\label{eq:candidate-u1}
    (q_0,q_1,q_2) = (m-3Q/2,Q,-\ell) \quad {\rm with} \quad h_\ell \le m < \ell/(p+r)\, , \ m\in\mathbb{Z}_{>0} \ ,
\end{align}
where $h_\ell$ denotes the conformal weight for the primary state whose 6d $U(1)$ electric charge is $\ell$. After the Higgsing to the rank-1 5d theory, the candidate primitive state reduces to a BPS state with mass
\begin{align}
    M_\ell \sim Qt_1 +(mp-\ell)t_2 \ .
\end{align}
Among these states, the one with maximum value of $(\ell-mp)$, if it exists, becomes a new K\"ahler cone generator. This occurs when $\ell -mp \approx \frac{Qb}{2p}$ under the assumption that $r$ is large. 

Therefore, it is possible that the 5d theory after the Higgsing contains a BPS state with mass proportional to $Q(t_1-\frac{b}{2p}t_2)$ at large $r$ and $p$. When this is the case, the actual K\"ahler cone would be formed by new K\"ahler parameters $\tilde{t}_1\equiv t_1-\frac{b}{2p}t_2$ and $t_2$. In other words, within the cone where $\tilde{t}_1,t_2\ge 0$, the electric charges $q_i$ for all BPS particle states, including those from 6d BPS strings, always satisfy $q_it_i\ge0$.
%
%
We can express the cubic prepotential using these new K\"ahler cone generators $\tilde{t}_1$ and $t_2$. Then the result is
\begin{align}
    6\mathcal{F} = 3p\tilde{t}_1^2 t_2 +9p^2 \tilde{t}_1t_2^2 + 9\left(-11p^3+36p^2r-39pr^2+18r^3-3r^4/p\right)t_2^3 \ .
\end{align}
We have numerically checked that this prepotential has positive CS coefficients and satisfies $\Delta_2\ge0$ for any sufficiently large $p,r$, with $r\ge p>r/2$.

We also examined other ranges of $p,r$ charges. The hypermultiplets that need to be flopped vary with each case. For example, in the charge range where $r/2\ge p >r/3$, the hypermultiplets with charges $(2,0,-r)$ and  $(3,0,-p-r)$ also need to be flopped, in addition to those already flopped when $p>r/2$.
However, when these flopped hypermultiplet contributions are carefully considered, we always find 5d prepotentials with positive CS coefficients and $\Delta_2\ge0$.

Likewise, we can explore other Higgsing possibilities using other KK states, such as those with charges $(1,0,-r)$ or $(1,0,-p-r)$. Although we will not provide detailed calculations here, our results indicate that these alternative Higgsing options tend to place less stringent constraints on the prepotentials of the Higgsed theories compared to the previous Higgsing using the charge $(1,0,-p)$ state.

Therefore, it seems that our moduli space analysis using the prepotential of the Higgsed 5d theory does not impose further constraints on the massless matter content of the infinite family of $U(1)$ theories, unless there is some other physical reason for the absence of candidate primitive states with small enough $m$ in \eqref{eq:candidate-u1}.

\paragraph{Models with 1-form symmetry} Nevertheless, we come across an interesting result for the infinite families with three charges $pn, rn,$ and $(p+r)n$, where $p$ and $r$ are co-prime. These theories have a $\mathbb{Z}_n$ 1-form symmetry in their massless spectra. As we will demonstrate, the Coulomb branch analysis can eliminate these theories when $n>1$. 

For these theories, we have a rather simple prepotential after a 5d reduction followed by the Higgsing process with a hypermultiplet of charge $(p,0,-pn)$ that we described earlier. We compute the prepotential of the resulting theory as
\begin{align}
    6\mathcal{F} &= 3nt_1^2t_2 - 9n^2\left(2p^2\!+\!2r^2\!+\!2pr\!-\!1\right)t_1t_2^2+9n^3\left(6p^3\!+\!6r^3\!+\!9p^2r\!+\!9pr^2\!-\!6p^2\!-\!6r^2\!-\!6pr\!+\!1\right)t_2^3 \nonumber \\
    &\quad + 54n^3\left(\sum_{i=1}^p(p-i)^3+\sum_{i=1}^r(r-i)^3+\sum_{i=1}^{p+r}(p+r-i)^3\right)t_2^3 \ ,
\end{align}
with $t_{1,2}$ defined in \eqref{eq:u1-t012}.
Here, the second line is the contributions from the flopped KK states from the 6d  hypermultiplets with charges $pn, rn$ and $(r+p)n$ that have $q_it_i<0$ along the limit for the Higgsing. Once again, the negative coefficient in the $t_1t_2^2$ term suggests that $t_1,t_2$ may not be the correct K\"ahler cone generators. 

We find that the conditions $\Delta_{1,2}\ge0$ and the positivity of CS coefficients for the prepotential within the K\"ahler cone can be satisfied only if there exist at least one primary state from the 6d BPS strings with
\begin{align}
    \ell = \tfrac{Qb}{n} = 6n(p^2+pr+r^2)Q \ , \quad m = \tfrac{Qb}{2n^2} = 3(p^2+pr+r^2)Q  \ , 
\end{align}
for some $Q$, where $m=h_\ell$ denotes the left-moving conformal dimension of the worldsheet primary state. If then this state gives rise to a 5d primitive BPS state whose dual K\"ahler parameter is proportional to $\tilde{t}_1=t_1 - 3n(p^2+pr+r^2) t_2$ after the Higgsing. The electric charges $q_i$ for all other BPS states satisfy $q_it_i\ge0$ when $\tilde{t}_1,t_2\ge0$. Therefore, the parameters $\tilde{t}_1,t_2$ become the generators of the K\"ahler cone in the Higgsed theory. Now, the prepotential written in terms of $\tilde{t}_1,t_2$ for any $p$ and $r$ is 
\begin{align}
    6\mathcal{F} &= 3n\tilde{t}_1^2 t_2 + 9n^2 \tilde{t}_1t_2^2 + 9n^3 t_2^3 \ .
\end{align}

This prepotential, for any $n$, satisfies the condition $\Delta_2 = (3n^2)^2 - 9n^3 \times n = 0$ at the boundary where $\tilde{t}_1\rightarrow 0$, and thus the Coulomb branch metric $M_{IJ}$ will have a vanishing eigenvalue with an eigenvector $(-3n,1)^T$ at that point. This signals the presence of a 5d rank-1 CFT at $\tilde{t}_1=0$. To further identify the CFT, we can rewrite the prepotential once more using the parameters $u_{1,2}$ defined as $\tilde{t}_1= - 3n u_1, t_2 = u_2+u_1$, which we deduce from the null vector of the metric $M_{IJ}$ at the boundary. The resulting prepotential then becomes
\begin{align}
    6\mathcal{F} = 9n^3u_1^3+9n^3u_2^3 \ .
\end{align}
Here we have assumed that the unit electric and magnetic charges for the 6d $U(1)$ gauge symmetry are quantized to be $\pm1$. The parameter $t_2$, thus also $u_1$, is normalized according to this quantization condition. The coefficient $9n^3$ of the $u_1^3$ term is the cubic CS coefficient for the local rank-1 CFT. However, since this CS coefficient cannot exceed `9', any theory with $n>1$ are inconsistent. Therefore, we conclude that all these theories with $n>1$ fall into the Swampland.

However, if the 6d $U(1)$ gauge theories have a $\mathbb{Z}_n$ 1-form symmetry, which matches the symmetry of the massless charge spectrum, and this 1-form symmetry is gauged, then the charge quantization condition changes. In this case, in order to read correct CS coefficients, we need to rescale the associated K\"ahler parameter as $t_2\rightarrow t_2/n$. With this rescaling, the prepotentials of the 5d Higgsed theories reduce to
\begin{align}
    6\mathcal{F}\quad \overset{t_2\rightarrow t_2/n}{\longrightarrow} \quad 6\mathcal{F} &= 3\tilde{t}_1^2 t_2 + 9 \tilde{t}_1t_2^2 + 9 t_2^3 \ .
\end{align}
This revised prepotential now aligns with the presence of a local rank-1 CFT, specifically the $E_0$ theory with CS coefficient `9', at the boundary $\tilde{t}_1\rightarrow0$.

In summary, our analysis of the K\"ahler moduli space indicates that the $U(1)$ theories with hypermultiplets carrying charges $pn, rn$, and $(p+r)n$, where $p,r$ are relatively prime, are all equivalent to the theory with hypermultiplets that have charges $p, r,$ and $p+r$: these theories must have a $\mathbb{Z}_n$ 1-form symmetry, and this symmetry must be gauged. This result shows concrete examples that are consistent with the Massless Charge Sufficiency Conjecture proposed in \cite{Morrison:2021wuv} and the No Global Symmetries Conjecture
in \cite{Banks:1988yz,Kallosh:1995hi,Banks:2010zn,Harlow:2018jwu,Harlow:2018tng,McNamara:2019rup,Harlow:2020bee}. 
The Massless Charge Sufficiency Conjecture states that the full charge lattice of a 6d supergravity theory coincides with the charge lattice generated solely by its massless spectrum.\footnote{More precisely, the conjecture for the massless charge sufficiency condition suggested in \cite{Morrison:2021wuv} holds only for gauge factors whose anomaly coefficients $b$ meet the condition $a\cdot b <0$. This conjecture is applicable to the cases discussed in this paper, as $a\cdot b<0$ for all $T=0$ theories.} Our analysis of the $U(1)$ theories supports this conjecture, as it shows that the $U(1)$ electric charge lattice must be generated by a unit charge $n={\rm GCD}(pn,rn,(p+r)n)$, the greatest common divisor of the electric charges in the massless spectrum of these theories.  Note, however, that such a conjecture cannot be generally true.  For example, the $Spin(32)/\mathbb{Z}_2$ heterotic strings provides a counterexample, where the massless fields do not generate the spinor representation, but the massive one does have it. Also, in 6d supergravity theories with anomaly coefficients $b$ that satisfy $a\cdot b \ge0$, massless fields are not always sufficient to generate the full charge lattice of gauge groups related to $b$ \cite{Morrison:2021wuv}. Perhaps the conjecture applies only to charge sectors that can potentially be protected by certain BPS conditions. Moreover, these theories now have a $\mathbb{Z}_n$ 1-form symmetry, which must be gauged to ensure that the prepotential meets the necessary consistency conditions. This is consistent with the absence of global symmetries in gravitational theories.

\subsection{$SU(2)$ theories}\label{sec:su2}

We now present a selection of potentially consistent anomaly-free 6d theories with $T=0$ and $SU(2)$ gauge algebras, and we examine their consistency by employing the analysis on the Coulomb branch moduli space of their 5d reductions and Higgsings. 

The first example is the $SU(2)$ theory with $n_{\bf 3}$ adjoint and $n_{\bf 2}$ fundamental hypermultiplets which we discussed in the cases with $1\le b\le 8$ in Sections \ref{sec:higgsing} and \ref{sec:chiralprimary}. We will now explore the theories for $9\le b\le 12$. It was shown in \cite{Raghuram:2020vxm} that the theories with $9\le b\le 11$ cannot be realized in F-theory. Any attempt to realize these theories in F-theory automatically results in symmetry enhancements to $(SU(2)\times U(1))/\mathbb{Z}_2$ for $b=9$, and to $(SU(2)\times SU(2))/\mathbb{Z}_2$ for $b=10,11$. On the other hand, the F-theory model for $b=12$ shows an $SO(3)=SU(2)/\mathbb{Z}_2$ gauge symmetry, rather than $SU(2)$ \cite{Morrison:2021wuv}. Thus, it is yet unclear whether these models with an $SU(2)$ gauge factor can consistently couple to gravity.

The prepotentials of these theories, after being compactified on a circle, are calculated in \eqref{eq:pre-su2-gravity}. Consider now a Higgsing with a vev of a BPS state with charge $q_0=1, q_2=-2$ which corresponds to a KK state from a 6d adjoint hypermultiplet. This Higgsing leads to a rank-1 theory with a prepotential, which can be obtained from \eqref{eq:pre-su2-gravity} by setting $t_0=0$, given by
\begin{align}
    6\mathcal{F}_{SU(2)_b}=6t_1^2t_2 +6(6-b)t_1t_2^2 + 2(6-b)^2 t_2^3 \ .
\end{align}
The presence of negative coefficients in the prepotential suggests that the theories with $9\le b\le 12$ must have additional primitive BPS states from their string spectra. Based on what we currently understand, there is no known method to clearly verify the existence of these states and their associated worldsheet primary fields. However, we can still deduce some intriguing conclusions as detailed below.

The BPS string carrying tensor charge $Q$ supports an $SU(2)$ current algebra at level $k=Qb$. The worldsheet primary field in spin $\ell/2$ representation of $SU(2)$, for $0< \ell\le Qb$, could yield candidate primitive states with electric charges
\begin{align}\label{eq:candidate-su2}
    (q_0,q_1,q_2) = (m-3Q/2,Q,-\ell) \quad {\rm with} \quad h_\ell \le m < \ell/q_{\rm max}\, , \ m\in\mathbb{Z}_{>0} \ ,
\end{align}
where $h_\ell$ denotes the conformal weight for the primary state, and $q_{\rm max}=2$ since the 6d theory includes adjoint fields.
After Higgsing to the rank-1 5d theory, the candidate primitive state reduces to a BPS state with mass
\begin{align}
    M_\ell \sim Qt_1 +(2m-\ell)t_2 \ .
\end{align}
If the worldsheet CFT fails to provide states with sufficiently small $m$, then the 6d theory cannot have a consistent prepotential in its 5d reduction.

For instance, the worldsheet CFT on a single string with $Q=1$ could have such primary fields that produce candidate primitive states with masses 
\begin{align}\label{eq:su2-primitive}
    b=9 \quad &: \quad \ell = 8 \ , \quad m=\lceil h_\ell \rceil = 2 \ ,  \quad M_\ell \sim \tilde{t}_1=t_1-4t_2 \ , \nonumber \\
    b=10 \quad &: \quad \ell = 10 \ , \quad m=\lceil h_\ell \rceil = 3 \ ,  \quad M_\ell \sim  \tilde{t}_1=t_1-4t_2 \ , \nonumber \\
    b=11 \quad &: \quad \ell = 11 \ , \quad m=\lceil h_\ell \rceil = 3 \ ,  \quad M_\ell \sim  \tilde{t}_1=t_1-5t_2 \ , \nonumber \\
    b=12 \quad &: \quad \ell = 12 \ , \quad m=\lceil h_\ell \rceil = 3 \ ,  \quad M_\ell \sim  \tilde{t}_1=t_1-6t_2 \ , 
\end{align}
respectively, after the Higgsing process. It is then possible that the K\"ahler cone on the Coulomb branch of the Higgsed theory is generated by $\tilde{t}_1$ and $t_2$ instead of $t_{1,2}$. If this is the case, we can rewrite the prepotential using $\tilde{t}_1,t_2$ as
\begin{align}\label{eq:cubic-b-9101112}
    6\mathcal{F}_{b=9} &= 6\tilde{t}_1^2t_2 + 30\tilde{t}_1t_2^2+42t_2^3 \ , \nonumber \\
    6\mathcal{F}_{b=10} &= 6\tilde{t}_1^2t_2 + 24\tilde{t}_1t_2^2+32t_2^3\ ,\nonumber \\
    6\mathcal{F}_{b=11} &= 6\tilde{t}_1^2t_2 + 30\tilde{t}_1t_2^2+50t_2^3 \ ,\nonumber \\
    6\mathcal{F}_{b=12} &= 6\tilde{t}_1^2t_2 + 36\tilde{t}_1t_2^2+72t_2^3 \ .
\end{align}

Now, the prepotential for the $b=9$ case satisfies $\Delta_{1,2}>0$ at the boundary $\tilde{t}_1\rightarrow 0$, so we cannot rule out the theory with $b=9$ based on our moduli space analysis. On the other hand, the prepotentials for the $b=10,11,12$ cases satisfy $\Delta_2=0$ at the boundary $\tilde{t}_1\rightarrow 0$, and thus they must have a local CFT at that point. Then, we can rewrite the prepotentials once again in terms of the K\"ahler parameters $u_1, u_2$, where $u_1$ is defined as the K\"ahler parameter for the rank-1 CFT. The results are
\begin{align}
    6\mathcal{F}_{b=10} &= 32u_1^3+32u_2^3\quad {\rm with} \quad \tilde{t}_1=-4u_1, \ t_2 = u_2 + u_1 \ ,\nonumber \\
    6\mathcal{F}_{b=11} &= 50u_1^3+50u_2^3 \quad {\rm with} \quad \tilde{t}_1=-5u_1, \ t_2 = u_2 + u_1\ ,\nonumber \\
    6\mathcal{F}_{b=12} &= 72u_1^3+72u_2^3 \quad {\rm with} \quad \tilde{t}_1=-6u_1, \ t_2 = u_2 + u_1\ .
\end{align}
Here, since the cubic CS coefficient for the CFT, which is the coefficient of the $u_1^3$ term, exceed the upper limit `9', these prepotentials for $b=10,11,12$ are all inconsistent.

However, it is too fast to conclude that the 6d theories for $b=10,11,12$ are inconsistent. There is still a possibility of finding other primitive states from the string spectra at higher tensor charges $Q>1$, whose electric charges cannot be expressed as a positive combination of charges of the primitive states we identified already. For instance, the theories with $b=10,11$ include such candidate primitive states with $\ell=40$ and $\ell=44$, respectively, from strings at $Q=4$. Then the prepotentials for $b=10,11$, when expressed in terms of the K\"ahler parameters associated with these states, show $\Delta_{1,2}>0$, and thus these theories cannot be easily eliminated. This also shows that  the analysis of just a single string spectrum is not enough to correctly identify the K\"ahler cone, and we may need to take into account the entire string spectrum for a complete analysis.

On the other hand, the worldsheet CFTs for $b=12$ cannot provide such additional primitive states. Taking into account all the candidate primitive states from strings at higher tensor charges, the K\"ahler cone of the 5d Higgsed theory is always bigger than or equal to the cone generated by $\tilde{t}_1$ and $t_2$ defined in \eqref{eq:su2-primitive}. In particular, when the K\"ahler cone is bigger than the cone of $\tilde{t}_1$ and $t_2$, the prepotential shows $\Delta_2 < 0$ and thus violates the metric positivity condition. Therefore, the theory for $b=12$, which has an $SU(2)$ gauge group, has an inconsistent prepotential in its 5d reduction and thus falls into the Swampland.

\paragraph{Models with 1-form symmetry} Recall that the $SU(2)$ theory for $b=12$ we are examining has a matter content of $n_{\bf 3}=55$ and $n_{\bf 2}=0$. It contains only adjoint hypermultiplets and thus has a $\mathbb{Z}_2$ 1-form symmetry in its massless spectrum. If this 1-form symmetry also applies to the massive spectrum and string excitations, and furthermore if it is gauged, meaning the gauge group is $SO(3)=SU(2)/\mathbb{Z}_2$ rather than $SU(2)$, the charge quantization changes and the K\"ahler parameter used in the prepotential needs to be rescaled accordingly. When this is the case, the minimum electric charge for the $U(1)\subset SU(2)$ gauge symmetry becomes $\pm2$, and the appropriate K\"ahler parameter for the $U(1)$ charge would require rescaling $t_2\rightarrow t_2/2$ to ensure that the Chern-Simons coefficients are properly quantized for the $SO(3)$ gauge group. With this rescaling, the cubic prepotential in \eqref{eq:cubic-b-9101112} can be written as
\begin{align}
    6\mathcal{F}_{b=12} = 3\tilde{t}_1^2t_2 + 9\tilde{t}_1t_2^2+9t_2^3 \ .
\end{align}
This result now shows the presence of a local CFT, the $E_0$ theory with CS coefficient `9', at the boundary $\tilde{t}_1=0$. Therefore, the theory for $b=12$ has a consistent prepotential if the gauge group is $SO(3)$ instead of $SU(2)$. This is consistent with the fact that the F-theory background for this theory has a $\mathbb{Z}_2$ torsion and thus the gauge group is in fact $SO(3)$, as discussed in \cite{Morrison:2021wuv}. Our analysis provides a concrete field-theoretic derivation supporting this result.

In fact, this analysis can be extended to other theories that have a 1-form symmetry in their massless spectra. Such theories with $T=0$ and an $SU(2)$ gauge symmetry occur with $b=4n$ for $n\ge 3$, and include hypermultiplets only in integral spin representations. Most of these theories lack F-theory realizations since they involves higher dimensional representations which cannot be realized in geometry. We can prove, as we will illustrate shortly, that all these theories have the same cubic prepotential, after a circle compactification and the same Higgsing process we previously described, given by
\begin{align}\label{eq:su2-pre-1-form}
    6\mathcal{F} = 6\tilde{t}_1^2t_2 + 36\tilde{t}_1t_2^2+72t_2^3 \ .
\end{align}
This occurs only if their string spectra include at least one primary field with $\ell=Qb$ and conformal dimension $m=h_\ell = Qb/4$. If this is not the case, the prepotentials would fail to meet the condition  $\Delta_2\ge0$, and thus the original 6d theory is inconsistent. 

Interestingly, this prepotential is identical to that of the theory with $b=12$. Thus, employing the same reasoning as before, we can conclude that all these theories must have a $\mathbb{Z}_2$ 1-form gauge symmetry extending beyond the massless spectra, and their gauge group must be $SO(3)=SU(2)/\mathbb{Z}_2$. This confirms the full charge lattice of these theories is generated by the massless charged states. 
Therefore, this result, which now covers all $SU(2)$ models with hypermultiplets in integral spin representations, along with the $U(1)$ infinite family examples investigated in the previous section, provides concrete and strong evidences supporting the Massless Charge Sufficiency Conjecture in \cite{Morrison:2021wuv}. 

As a particular example to demonstrate this analysis, let us examine a 6d $SU(2)$ theory with $b=40$ that includes matter content given by
\begin{align}
    n_{\bf 3}=48 \,, \quad n_{\bf 5}=21 \,, \quad n_{\bf 7} = 2 \ ,
\end{align}
where $n_{\bf 2j+1}$ denotes the number of hypermultiplets in spin $j$ representation, which has no known F-theory construction. Using the Lagrangian in \eqref{eq:Lagrangian}, we compute the prepotential of the 5d theory after a circle reduction as
\begin{align}
    6\mathcal{F} = 9t_0^3+9t_0^2t_1+3t_0t_1^2+162t_0^2t_2+18t_1^2t_2+252t_0t_2^2+84t_1t_2^2+108t_0t_1t_2+88t_2^3 \ ,
\end{align}
where $t_0, t_1$, and $t_2$ represent the K\"ahler parameters for 5d BPS states with electric charges $(q_0,q_1,q_2)=(1,0,-6), (-3/2,1,0),$ and $(0,0,2)$ respectively. 

Then we proceed to Higgs this 5d theory by giving a vev to a KK hypermultiplet with charge $(1,0,-2)$. This is achieved at the level of K\"ahler parameters by setting $t_0=-4t_2$. In this setting, the K\"ahler parameters for the KK hypermultiplets with charges $(1,0,-4), (1,0,-6),$ and $(2,0,-6)$ become negative, and thus flop transitions for these hypermultiplets will take place. Then, the prepotential of the Higgsed theory is given by
\begin{align}
    6\mathcal{F} =&\ 6t_1^2t_2-204t_1t_2^2+1096t_2^3 + 2(8n_{\bf 5}+(64\!+\!8\!+\!8)n_{\bf 7})t_2^3 \nonumber\\
    =&\ 6t_1^2t_2-204t_1t_2^2+1752t_2^3 \ .
\end{align}
Here, the terms with $n_{\bf 5}$ and $n_{\bf 7}$ correspond to the flop transition factors.

This prepotential can satisfy the consistency condition $\Delta_2\ge0$ only if the string spectra contain at least one primary field with $\ell=40Q$ and conformal dimension $m=h_\ell = 10Q$. Assuming such a primary field exists, and by using the K\"ahler cone generators $\tilde{t}_1\equiv t_1-20t_2$ and $t_2$ in this case, the prepotential can finally be recast as that in \eqref{eq:su2-pre-1-form}. This shows that this 6d theory can have a consistent prepotential only if the theory has a $\mathbb{Z}_2$ 1-form gauge symmetry and thus the gauge group is $SO(3)$.

The same computation applies to all other $SU(2)$ models that have $n_{\bf 2j+1}$ hypermultiplets in $SU(2)$ spin $j\in\mathbb{Z}$ representations. After the Higgsing with a hypermultiplet that has charge $(1,0,-2)$, the prepotential simplifies to
\begin{align}
    6\mathcal{F} =&\ 6\tilde{t}_1^2t_2+36\tilde{t}_1t_2^2+(64-18b-\tfrac{3}{2}b^2)t_2^3 + \sum_{j=1}^\infty 2j^2(j+1)^2n_{\bf 2j+1}t_2^3 \nonumber \\
    &\ +\sum_{j=1}^\infty\frac{4}{15}j(j^2-1)(3j^2-2)n_{\bf 2j+1}t_2^3\nonumber\\
    =&\ 6\tilde{t}_1^2t_2+36\tilde{t}_1t_2^2+(64-18b-\tfrac{3}{2}b^2)t_2^3 + \sum_{j=1}^\infty \frac{2}{15}j(j+1)(2j+1)(3j^2+3j+4)n_{\bf 2j+1}t_2^3 \nonumber \\
    =&\ 6\tilde{t}_1^2t_2+36\tilde{t}_1t_2^2+72t_2^3 \ ,
\end{align}
with the K\"ahler cone generator $\tilde{t}_1 \equiv t_1-b/2t_2$ associated to a primary field with $\ell=Qb$ and conformal dimension $m=h_\ell = Qb/4$. The summation over $j$ in the first line represents contributions to the $t_2^3$ term from the 6d hypermultiplets given in \eqref{eq:pre-pert} with $\sum_{i=1}^j(2i)^3=2j^2(j+1)^2$ for a spin $j$ representation, and the summation in the second line corresponds to contributions from the flopped hypermultiplets, which is $\sum_{i=1}^j\sum_{l=1}^{i-1}(2i-2l)^3$ for a spin $j$ rep with KK charge $l$, during we take the Higgsing process. In the last line, we use the definitions of the anomaly coefficients $a$ and $b$ in \eqref{eq:anomaly-cancel}, along with the $SU(2)$ group theory factors
\begin{align}
    A_{\bf 2j+1} = \frac{2j(j+1)(2j+1)}{3} \ , \quad C_{\bf 2j+1} = \frac{j(j+1)(2j+1)(3j^2+3j-1)}{15} \ .
\end{align}
These elements together lead to
\begin{align}
    6a\cdot b-3/2b^2 =& \  A_{\bf adj}-\sum_{j}n_{\bf 2j+1}A_{\bf 2j+1}-\frac{1}{2}\left(\sum_{j}n_{\bf 2j+1}C_{\bf 2j+1}-C_{\bf adj}\right)\nonumber \\
    =& \ 8 - \sum_{j=1}^\infty \frac{2}{15}j(j+1)(2j+1)(3j^2+3j+4)n_{\bf 2j+1} \ ,
\end{align}
where `8' in the second line comes from the $SU(2)$ vector multiplet. This proves that  after undergoing 5d reductions and Higgsings, all these theories have the same cubic prepotential. Consequently, as previously mentioned, we can conclude that the global form of their gauge group is $SO(3) = SU(2)/\mathbb{Z}_2$. If not, the theories are inconsistent.

This analysis also yields strong predictions for both the BPS states and the worldsheet primary states in the $SU(2)$ models that exhibit a $\mathbb{Z}_2$ 1-form symmetry. In fact, worldsheet primary fields with $\ell=Qb$ and $h_\ell=Qb/4$ are mapped to BPS states with electric charge $3Q$ in the $5d$ $E_0$ theory embedded in the Coulomb branch of the 5d Higgsed theory. Therefore, these primary fields should exist for every tensor charge $Q$, and moreover their degeneracies (or spin contents) should  agree with those of BPS states in the $E_0$ theory. For instance, the BPS invariants (or genus-0 GV-invariants) in the $E_0$ theory, such as $3,-6,27,\cdots$ for the states with electric charges $3,6,27,\cdots$, provide predictions for the spin contents of the worldsheet primary states in the $SO(3)$ models. 

\section{Conclusions and future directions}\label{sec:5}

In this paper, we have investigated consistency conditions on the cubic Chern-Simons coefficients in the 5d effective Lagrangians of 6d $\mathcal{N}=(1,0)$ supergravity theories compactified on a circle, based on the positivity of the metric and string tensions in their Coulomb branch moduli space. These conditions hold within the K\"ahler cone, where the central charges of all BPS particle states are non-negative.
We proposed a method to precisely determine the K\"ahler cone by identifying primitive BPS states, which span the electric charge lattice of 5d BPS states, that arise from 6d massless matter fields, their KK momentum modes, and worldsheet primary fields on 6d BPS strings. In particular, additional constraints on the 5d effective Lagrangians are derived from the consistency of Higgs branch renormalization group (RG) flows and the potential emergence of local 5d SCFTs at specific points on the Coulomb branch.

Using these consistency criteria, we have examined a broad range of potentially consistent anomaly-free 6d supergravity theories with $T=0$ and a rank-1 gauge algebra. Surprisingly, although these theories largely lack UV completions in F-theory or string theory, most of them  can meet the constraints we found, provided certain primary states exist in their BPS string spectra. We are not able to disprove the existence of these primary states with current knowledge. However, for a specific subset of theories that feature non-trivial discrete 1-form symmetry in their massless spectra, we can definitively identify such primary states and demonstrate that the 1-form symmetry must be the symmetry of the full theory and also gauged. This analysis assumes a complete classification of 5d rank-1 SCFTs, which is crucial to our conclusions. Specifically, we have demonstrated that infinite families of $U(1)$ gauge theories, which include three types of hypermultiplets with charges $pn, rn,$ and $(p+r)n$ for $n>1$ and co-primes $p,r$, and $SU(2)$ gauge theories that contain hypermultiplets only in integral spin representations, actually fall into the Swampland unless their discrete 1-form symmetry, such as $\mathbb{Z}_n$ for $U(1)$ models or $\mathbb{Z}_2$ for $SU(2)$ models, is gauged. These results provide strong field-theoretical evidence supporting the Massless Charge Sufficiency Conjecture proposed in \cite{Morrison:2021wuv} as well as the No Global Symmetries Conjecture in \cite{Banks:1988yz,Kallosh:1995hi,Banks:2010zn,Harlow:2018jwu,Harlow:2018tng,McNamara:2019rup,Harlow:2020bee}.

A natural extension of this study is to explore constraints on the moduli spaces of other anomaly-free 6d theories that include more tensors and higher rank gauge algebras. This requires more systematic investigations into the positivity conditions of CS coefficients in the 5d effective Lagrangian, as these conditions become increasingly complex with the addition of more tensor or vector multiplets. Moreover, acquiring more data on higher rank 5d SCFTs, such as higher rank CFT classifications, would be essential. Such higher rank studies could also yield further constraints on the 6d rank-2 theories that we examined in this work.
Remember that we applied the metric positivity conditions solely to the Higgsed rank-1 theory because it contains fewer CS coefficients and is easier to analyze than the original theory. Thus, these conditions represent only a subset of the positivity conditions in the original rank-2 theory before Higgsing, and it would be interesting to check whether the full set of positivity conditions can be used to rule out other rank-2 gravity theories.

Extending this work to higher rank theories could involve proving the Massless Charge Sufficiency Conjecture for $T=0$ theories with other non-Ablian gauge algebras that have non-trivial 1-form symmetries in their massless matter content.
Also, as observed in our proof for the rank-1 gauge theories, the conjecture appears to be deeply connected to hidden relationships between group theoretic factors such as $A_{\bf r}^G$ and $C_{\bf r}^G$ for particular representations ${\bf r}$, and the cubic indices that we used to compute 1-loop contributions to the CS coefficients from the 6d massless fields and 5d flopped hypers. Employing similar Higgsings that we performed with non-trivial KK states, combined with these relationships, may enable us to prove the conjecture for general non-Abelian gauge algebras.

Lastly, the most critical part of our analysis is to know the exact K\"ahler cone. To achieve this, it is crucial to identify the precise spectrum of primary fields in 2d $(0,4)$ worldsheet CFTs on 1/2 BPS strings. The current algebra in these 2d CFTs allows us to establish lower bounds on the conformal dimensions of primary fields for given charges, and we can use this to propose candidate primitive BPS states associated with those primary fields. However, this information alone was not enough to conclusively identify the precise boundaries of the K\"ahler cone and, consequently, to determine the consistency of a 6d theory. Additional data on worldsheet primary fields, possibly derived from modular properties or $(0,4)$ superconformal algebras, is necessary to reach a definitive conclusion.

Conversely, we can extract non-trivial spectrum information on BPS strings from the consistency constraints on the moduli space of gravitational theories, as demonstrated by our identification of the existence and spin content of certain worldsheet primary fields in the $SU(2)$ models. The interplay between worldsheet data and Coulomb branch analysis could provide profound insights into both 2d CFTs and 6d supergravity.

\bigskip

\acknowledgments
We would like to thank Naomi Gendler, Minsung Kim, Houri Tarazi and Kai Xu for useful discussions.  H.K. is supported by Samsung Science and Technology Foundation under Project Number SSTF-BA2002-05 and by the National Research Foundation of Korea (NRF) grant funded by the Korea government (MSIT) (2023R1A2C1006542). The work of H.K. at Harvard University is supported in part by the Bershadsky Distinguished Visiting Fellowship. C.V. is supported in part by a grant from the Simons Foundation (602883,CV), the DellaPietra Foundation, and by the NSF grant PHY-2013858.

\bibliographystyle{JHEP}
\bibliography{refs}
\end{document}